\documentclass[journal]{IEEEtran}

\usepackage{amsmath}
\usepackage{amssymb}
\usepackage{amsfonts}
\usepackage{hyperref}
\usepackage[plain]{fancyref}
\usepackage[]{graphicx} 
\usepackage{diagbox}
\usepackage{mathrsfs}
\usepackage[version=3]{mhchem}
\usepackage{algorithm}
\usepackage[noend]{algpseudocode}
\usepackage{varwidth}
\usepackage{parnotes}
\renewcommand*\parnoteintercmd[1]{,\,}
\usepackage[multiple]{footmisc}
\usepackage{xcolor}
\usepackage{textcomp}
\RequirePackage{lineno}
\usepackage{cite}

\makeatletter
\def\ps@IEEEtitlepagestyle{%
  \def\@oddfoot{\mycopyrightnotice}%
  \def\@evenfoot{}%
}
\def\mycopyrightnotice{%
  \begin{centering}
    \begin{minipage}[b][12pt][t]{0.85\textwidth}{
      \begin{center}
      {\footnotesize \copyright 2020 IEEE. Personal use of this material is permitted. Permission from IEEE must be obtained for all other uses, in any current or future media, including reprinting/republishing this material for advertising or promotional purposes, creating new collective works, for resale or redistribution to servers or lists, or reuse of any copyrighted component of this work in other works.}
      \end{center}
    }
    \end{minipage}
  \end{centering}
}
\makeatother

\graphicspath{{./fig/}}

\begin{document}

\title{Reconstructing the Position and Intensity of Multiple Gamma-Ray Point Sources with a Sparse Parametric Algorithm}

\author{J.\,R.~Vavrek, D.~Hellfeld, M.\,S.~Bandstra, V.~Negut, K.~Meehan, W.\,J.~Vanderlip, J.\,W.~Cates, R.~Pavlovsky, B.\,J.~Quiter, R.\,J.~Cooper, T.\,H.\,Y.~Joshi%

\thanks{Manuscript received 5 August 2020. This material is based upon work supported by the Defense Threat Reduction Agency under HDTRA 10027-28018 \& 10027-30529. This support does not constitute an express or implied endorsement on the part of the United States Government. Distribution A: approved for public release, distribution is unlimited.

This document was prepared as an account of work sponsored by the United States Government. While this document is believed to contain correct information, neither the United States Government nor any agency thereof, nor the Regents of the University of California, nor any of their employees, makes any warranty, express or implied, or assumes any legal responsibility for the accuracy, completeness, or usefulness of any information, apparatus, product, or process disclosed, or represents that its use would not infringe privately owned rights. Reference herein to any specific commercial product, process, or service by its trade name, trademark, manufacturer, or otherwise, does not necessarily constitute or imply its endorsement, recommendation, or favoring by the United States Government or any agency thereof, or the Regents of the University of California. The views and opinions of authors expressed herein do not necessarily state or reflect those of the United States Government or any agency thereof or the Regents of the University of California.

This manuscript has been authored by an author at Lawrence Berkeley National Laboratory under Contract No.~DE-AC02-05CH11231 with the U.S.~Department of Energy. The U.S.~Government retains, and the publisher, by accepting the article for publication, acknowledges, that the U.S.~Government retains a non-exclusive, paid-up, irrevocable, world-wide license to publish or reproduce the published form of this manuscript, or allow others to do so, for U.S.~Government purposes.}%
\thanks{J.\,R.~Vavrek, D.~Hellfeld, M.\,S.~Bandstra, V.~Negut, K.~Meehan, J.\,W.~Cates, R.~Pavlovsky, B.\,J.~Quiter, R.\,J.~Cooper, and T.\,H.\,Y.~Joshi are with the Applied Nuclear Physics Program at Lawrence Berkeley National Laboratory, Berkeley, CA 94720 USA (e-mail: \href{mailto:thjoshi@lbl.gov}{thjoshi@lbl.gov}).}%
\thanks{W.\,J.~Vanderlip is with the Nuclear Engineering Department at the University of California, Berkeley, Berkeley, CA 94720 USA.}%
}

\maketitle

\markboth{IEEE Transactions on Nuclear Science}%
{---}

\newcommand{\nll}{\ell(\boldsymbol{x} | \hat{\boldsymbol{\lambda}})}

\begin{abstract}
We present an experimental demonstration of Additive Point Source Localization (APSL), a sparse parametric imaging algorithm that reconstructs the 3D positions and activities of multiple gamma-ray point sources.
Using a handheld gamma-ray detector array and up to four $8$~{\textmu}Ci $^{137}$Cs gamma-ray sources, we performed both source-search and source-separation experiments in an indoor laboratory environment.
In the majority of the source-search measurements, APSL reconstructed the correct number of sources with position accuracies of ${\sim}20$~cm and activity accuracies (unsigned) of ${\sim}20\%$, given measurement times of two to three minutes and distances of closest approach (to any source) of ${\sim}20$~cm.
In source-separation measurements where the detector could be moved freely about the environment, APSL was able to resolve two sources separated by $75$~cm or more given only ${\sim}60$~s of measurement time.
In these source-separation measurements, APSL produced larger total activity errors of ${\sim}40\%$, but obtained source separation distances accurate to within $15$~cm.
We also compare our APSL results against traditional Maximum Likelihood-Expectation Maximization (ML-EM) reconstructions, and demonstrate improved image accuracy and interpretability using APSL over ML-EM.
These results indicate that APSL is capable of accurately reconstructing gamma-ray source positions and activities using measurements from existing detector hardware.
\end{abstract}
\section{Introduction}\label{sec:intro}

\IEEEPARstart{G}{amma-ray} imaging is the inverse problem that aims to reconstruct the source term (both spatial and intensity) of gamma- or hard X-ray photons in an environment from measurements of photon counts.
Configurations of static detectors observing a stationary volume are typically used in medical imaging, whereas in geological mapping and nuclear security, measurements are often made using one or several detectors moving through the environment.
Regardless of application, the pose (i.e., position and orientation) of the gamma-ray detectors must be known in order to properly attribute source activity to different locations in the environment.
In the dynamic case, this detector pose information may be provided by coupling the gamma-ray detectors to a global positioning system (GPS)~\cite{lyons2012aerial, tansey2017multiscale} or inertial measurement unit (IMU)~\cite{cooper2013integration, joshi2017measurement}, or by using simultaneous localization and mapping (SLAM)~\cite{Durrant-Whyte2006_1, Durrant-Whyte2006_2, Hess2016, haefner2017handheld, Pavlovsky2019} or related methods~\cite{mascarich2018radiation, lee2018volumetric}.

Traditionally, the inversion problem may be solved by discretizing the spatial dimensions and employing some variant of maximum likelihood parameter estimation~\cite{Miller2015} or the Maximum Likelihood-Expectation Maximization (ML-EM) algorithm~\cite{Shepp1982}.
A common extension of ML-EM replaces the maximum likelihood with a maximum \textit{a posteriori} (MAP-EM) formulation, adding regularization or prior terms to the likelihood function in order to impose assumptions on the source distribution.
While the generality of these formulations enables use in a wide variety of scenarios, ML-EM and MAP-EM can be susceptible to overfitting (especially in noisy and underdetermined scenarios)~\cite{Reader2020, Bissantz2008}, and have resolutions limited by the discretization of spatial coordinates.

In previous work~\cite{Hellfeld2019}, we proposed Additive Point Source Localization (APSL), a sparse parametric image reconstruction algorithm, as an alternative to ML-EM, MAP-EM, and several other previous methods~\cite{rao2008identification, deb2011radioactive, deb2013iterative, Bhattacharyya2018, Miller2015, Ristic2010, Chin2008, Vilim2009, Sharma2016, Cordone2017}.
APSL is proposed for sparse 3D scenarios with multiple point sources and unknown backgrounds, a situation where previous methods may have limited utility due to algorithmic assumptions.
In APSL, the image is considered the sum of radioactive point sources whose position and intensity ($\vec{r}_s, w_s$) are continuous in nature.
APSL adds one source at a time, comparing each model iteration with a statistically-founded stopping criterion in order to mitigate over-fitting.
In simulated measurements, the inherent point-source assumption and continuous variables yielded images with substantially improved accuracy and interpretability as compared with ML-EM or MAP-EM.

In this work, we demonstrate APSL using experimental data from a handheld gamma-ray detector system.
Section~\ref{sec:methods} begins with the Poisson likelihood formulation of gamma-ray imaging, and describes the APSL algorithm in terms of minimizing the Poisson negative log-likelihood and applying model selection criteria.
Section~\ref{sec:expdesign} covers the detector system used, the source-search and source-separation measurements performed, and further aspects of the reconstruction analysis.
Section~\ref{sec:results} presents the source-search and source-separation results, and evaluates the reconstruction performance of APSL against both ML-EM reconstructions and ground-truth source positions and activities.
Section~\ref{sec:discussion} then concludes with a discussion of APSL vs ML-EM, systematic uncertainties, and possible future work.

\section{Methods}\label{sec:methods}

Poisson statistics govern gamma-ray measurements.
A set of $I$ measurements of gamma-ray counts $\boldsymbol{x}$ may be modeled as Poisson random samples from a set of mean generative values~$\boldsymbol{\lambda}$:
\begin{align}\label{eq:poisson_model}
    \boldsymbol{x} \sim \text{Poisson}(\boldsymbol{\lambda}),\text{ with } \boldsymbol{\lambda} \equiv \boldsymbol{v}_{s} w_s + b\boldsymbol{t},
\end{align}
where $\boldsymbol{v}_s$ is the vector of system responses (often obtained through modeling or experiment) describing the sensitivity of each measurement $i$ to a single point source of activity $w_s$ at position $\vec{r}_s$, and $\boldsymbol{t}$ is the vector of integration times for each measurement.
The background rate $b$ may be known from dedicated background measurements and may in fact vary with time; in this work, $b$ is left as a constant free parameter to be determined (for a given detector).
The $i^\text{th}$ component of the system response (neglecting attenuation from the air or intervening objects) is in turn
\begin{align}\label{eq:sysresp}
    v_{si} \simeq \frac{\eta(\vec{r}_s, \vec{r}_i, \boldsymbol{q}_i) t_i}{4\pi \Tilde{r}_{si}^2},
\end{align}
where $\eta(\vec{r}_s, \vec{r}_i, \boldsymbol{q}_i)$ is the effective area of a detector\footnote{The effective area $\eta$ is defined as the conversion factor from a flux density $\varphi$ with dimensions of particles per unit area per unit time to a detected rate $R$ with dimensions of counts per unit time: $R = \eta \varphi$.
Given instead a flux $\phi$ with dimensions of particles per unit time, we can also write $R = \epsilon_\text{int} \epsilon_\text{geom}\phi$, where $\epsilon_\text{geom}$ and $\epsilon_\text{int}$ are the geometric and intrinsic detector efficiencies.
Equating the two expressions for $R$ and making the isotropic point source assumption that $\varphi = \phi / (4\pi r^2)$, where $r$ is the source-to-detector distance, we see that $\eta = \epsilon_\text{int} \epsilon_\text{geom} 4\pi r^2$.
For a far-field point source, $\epsilon_\text{geom} = A_\text{det}/(4\pi r^2)$, where $A_\text{det}$ is the area of the detector exposed to the flux, in which case $\eta = A_\text{det} \epsilon_\text{int}$, hence the name `effective area.'
Note that the definition of $v_i$ and thus $\eta$ differ by a factor of $4\pi$ from Ref.~\cite{Hellfeld2019}.
}
at position $\vec{r}_i$ and orientation $\boldsymbol{q}_i$ for a point source at $\vec{r}_s$, and $\Tilde{r}_{si}$ is a regularized version of the source-to-detector distance $r_{si} = \vert \vec{r}_s - \vec{r_i} \vert$.
The regularization corrects the usual $1/r^2$ intensity scaling for near field-effects in the non-zero size $d \simeq 5$~cm of the detector, and is given by
\begin{align}\label{eq:regularization}
    \frac{1}{\Tilde{r}_{si}^2} = \frac{r_{si}^2}{r_{si}^4 + d^4}.
\end{align}

In this formulation, the negative log-likelihood of observing the counts $\boldsymbol{x}$ given the set of Poisson mean values $\boldsymbol{\lambda}$ is
\begin{align}
    \ell(\boldsymbol{x} | \boldsymbol{\lambda}) = \left[ \boldsymbol{\lambda} - \boldsymbol{x} \odot \log\boldsymbol{\lambda} + \log\Gamma(\boldsymbol{x}+1) \right]^\intercal \cdot \boldsymbol{1}
\end{align}
where $\odot$ denotes element-wise multiplication.
Maximum likelihood estimation of the unknown parameter set $\boldsymbol{S} = \{ w_s, \vec{r}_s, b \}$ can now be formulated as an optimization problem continuous in both the intensity $w_s$ and 3D spatial coordinates $\vec{r}_s$ of a point source, as well as the unknown background $b$:
\begin{align} \label{psloptimization}
\boldsymbol{\hat{S}} = \underset{{(w_s, \vec{r}_s, b)}}{\operatorname{argmin}} ~\ell(\boldsymbol{x} | \boldsymbol{\lambda}).
\end{align}
Though solving for the maximum likelihood estimates of $w_s$ and $\vec{r}_s$ simultaneously is non-convex, this formulation drastically reduces the number of free variables considered in the optimization problem compared to, e.g., ML-EM techniques that solve for a source distribution $w(x,y,z)$ over an entire voxelized space.
As discussed in Ref.~\cite{Hellfeld2019}, we adopt a hybrid optimization approach: a non-convex optimization is run over $\vec{r}_s$ space, and at each trial $\vec{r}_s$, the optimum $w_s$ and $b$ are determined using ML-EM.

Moreover, the additive nature of Poisson random variables facilitates the inclusion of constant contributions from $K$ known sources with activities $w_k$:
\begin{align}\label{eq:poisson_model_multi}
    \boldsymbol{\lambda} \equiv \boldsymbol{v}_s w_s + b\boldsymbol{t} + \sum_{k=1}^{K} \boldsymbol{v}_k w_k
\end{align}
With this redefinition, (\ref{psloptimization}) reconstructs the unknown source parameters $(w_s, \vec{r}_s)$ and background $b$ in the presence of known source parameters $w_k$ and $\vec{r}_k$, $k=1,\ldots,K$.
This fact suggests the iterative Algorithm~\ref{apslalgo}~\cite{Hellfeld2019} for this sparse inverse problem.
Sources are added to the optimization problem one by one, treating previously-added sources as known terms $w_k$ and $\vec{r}_k$.
New source parameters $w_s$ and $\vec{r}_s$ (plus the background $b$) are then reconstructed via non-convex optimization, holding the $w_k$ and $\vec{r}_k$ fixed.
After each new source is identified, all parameters $\boldsymbol{S} \equiv \{(w_1, \vec{r}_1), \ldots, (w_K, \vec{r}_K), (w_w, \vec{r}_s); b \}$ are re-optimized simultaneously, again using the hybrid approach discussed above of optimizing positions in optimal weight space.
This re-optimization is typically the most computationally expensive step, often taking ${\sim}60\%$ of the total algorithm runtime.
Several cleaning procedures are then applied to combine spatially-close sources and remove sources that contribute weakly to the model counts, activity, or reduction in Bayesian Information Criteria (BIC)~\cite{Schwarz1978}.
The BIC is proportional to the negative log-likelihood $\ell(\boldsymbol{x} | \boldsymbol{\hat{\lambda}})$ evaluated using the optimal parameters $\boldsymbol{\hat{\lambda}} \equiv \boldsymbol{\lambda}(\boldsymbol{\hat{S}})$, but includes an additive penalty term on the number of model parameters.
The model with the lowest BIC is thus the preferred model.
The BIC of the $n$-source, single-detector APSL model for the set of $I$ measurements can be written as
\begin{align}
    \text{BIC} = 2 \nll + (4n+1) \log(I)
\end{align}
and provides the primary stopping criterion: if the BIC of the $(n+1)$-source model exceeds that of the $n$-source model, the $(n+1)$-source model is rejected.
As a secondary stopping criterion, we estimate the $p$-value of each model from the Poisson deviance between the model-predicted expected counts $\boldsymbol{\hat{\lambda}}$ and the measured counts $\boldsymbol{x}$, and select the model if $p>0.05$.

\begin{algorithm}[htbp]
\footnotesize
\caption{\footnotesize Additive Point-Source Localization~\cite{Hellfeld2019}}
\label{apslalgo}
\begin{algorithmic}[1]
\State Initialize background-only model $\boldsymbol{S} = \{(); b = \textrm{median}(\boldsymbol{x})\}$
\State converged = False
\If{$p > 0.05$}
    \State converged = True
\EndIf
\While{not converged}
    \State $\boldsymbol{S}_{\textrm{old}}=\boldsymbol{S}$
    \State Solve (\ref{psloptimization}) for new source with $\boldsymbol{\lambda}$ from (\ref{eq:poisson_model}) or (\ref{eq:poisson_model_multi}), append to $\boldsymbol{S}$
    \State Re-optimize source positions, intensities and backgrounds; update $\boldsymbol{S}$
    \State Test for acceptance of $\boldsymbol{S}$ relative to $\boldsymbol{S}_{\textrm{old}}$ using BIC
    \If{accepted}
        \State \begin{varwidth}[t]{\linewidth-2em}
        Drop weakly contributing sources and collapse nearby sources
        \end{varwidth}
        \State \begin{varwidth}[t]{\linewidth-2em}
        Re-optimize source positions, intensities, backgrounds; update $\boldsymbol{S}$
        \end{varwidth}
        \If{$p > 0.05$}
            \State converged = True
        \EndIf
    \Else
        \State $\boldsymbol{S} = \boldsymbol{S}_{\textrm{old}}$
        \State converged = True
    \EndIf
\EndWhile
\end{algorithmic}
\end{algorithm}

\section{Experimental design} \label{sec:expdesign}
\subsection{Detection system}
\begin{figure}[!htbp]
    \begin{center}
    \resizebox{.99\columnwidth}{!}{%
        \includegraphics[height=3cm]{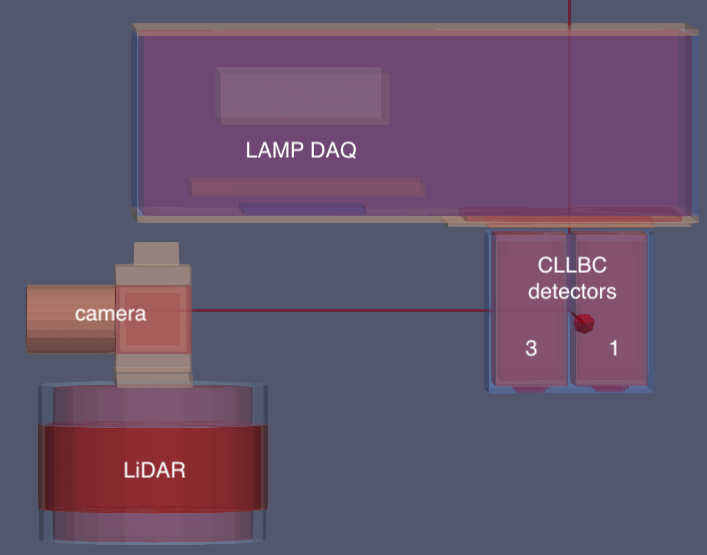}%
        \quad
        \includegraphics[height=3cm]{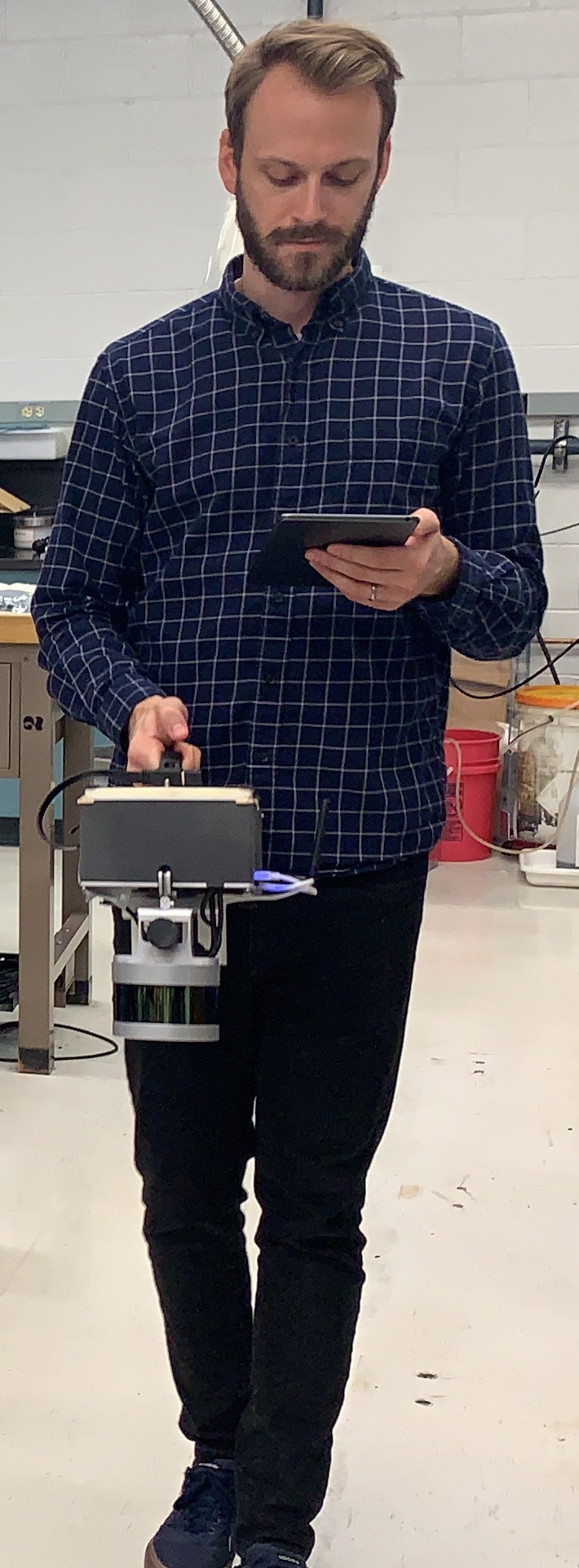}%
    }
    \end{center}
    \caption{Left: semi-transparent side view render of the NG-LAMP~\cite{Pavlovsky2018} detector system.
    Detector crystals~3 and 1 are visible in the foreground, and obscure crystals 2 and 0, respectively.
    Right: Photo of NG-LAMP (front view) carried by a human operator.}
    \label{fig:ng_render}
\end{figure}

APSL experiments were performed using NG-LAMP~\cite{Pavlovsky2019}, a $2\times 2$ array of CLLBC scintillator detectors (manufactured by Radiation Monitoring Devices Inc., total volume $130$~cm$^3$) read out by a handheld Localization and Mapping Platform (LAMP) electronics package developed at Lawrence Berkeley National Laboratory (LBNL)~\cite{Pavlovsky2018}.
The LAMP system collects synchronized gamma-ray, LiDAR, IMU, and video data.
The $2 \times 2$ crystal array produces an active-masked non-isotropic angular response function, providing directionality in a similar fashion as a passive coded aperture but without the loss of photons attenuated by the mask.
Attenuation from other LAMP components such as the LiDAR and data acquisition system (DAQ) contributes additional anisotropy to the response function---see Fig.~\ref{fig:ng_render}.

\begin{figure}[!htbp]
    \centering
    \includegraphics[width=\columnwidth]{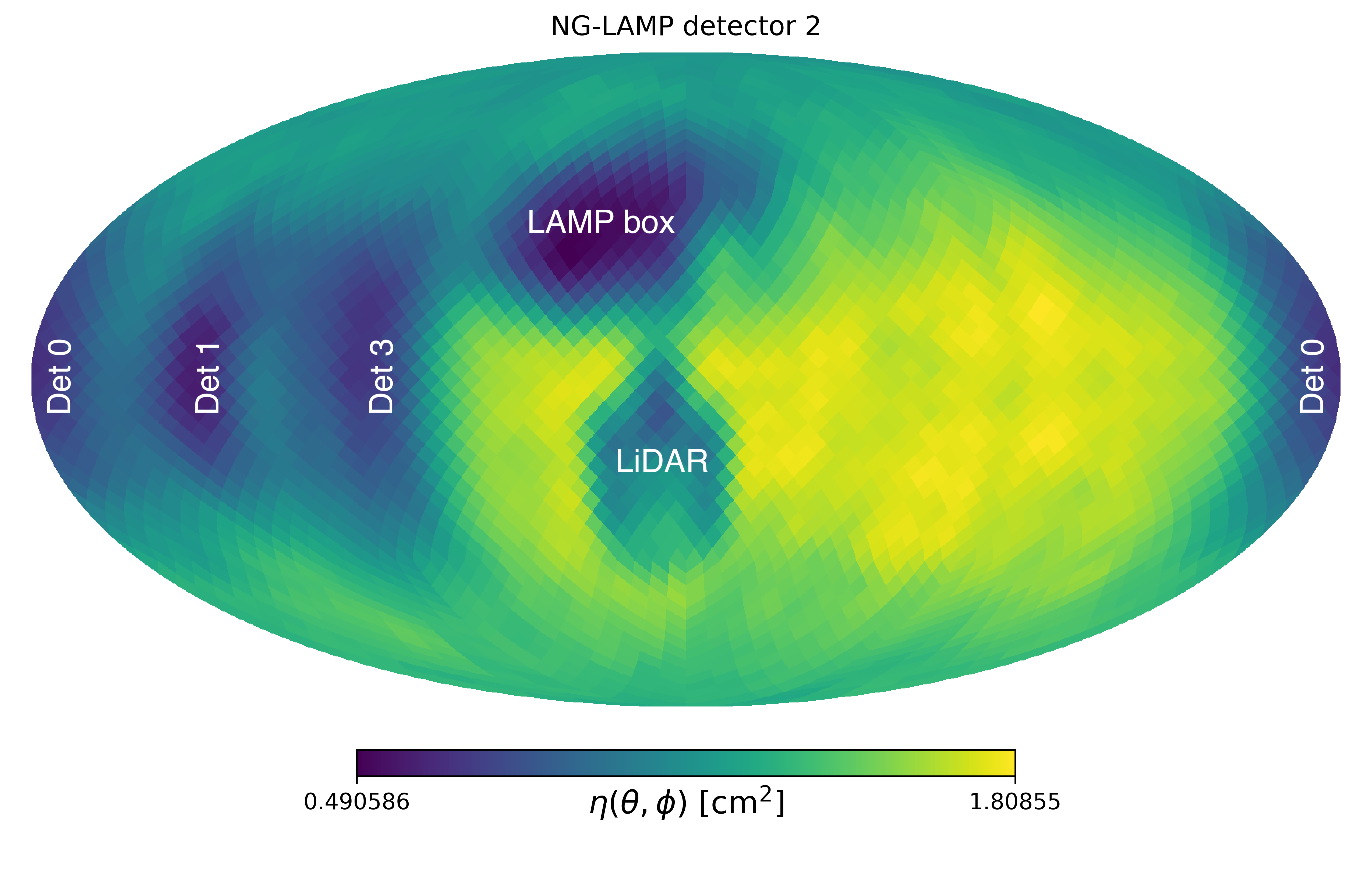}
    \caption{Mollweide projection of the $4\pi$ angular response function at $662$~keV for detector~2 in NG-LAMP (see Fig.~\ref{fig:ng_render}), given as the effective area $\eta$ for detecting a photopeak count.}
    \label{fig:ang_resp_2}
\end{figure}

In previous simulation studies~\cite{Hellfeld2019}, only isotropic detector response functions $\eta = \text{const}$ were considered.
In this work, we consider the full anisotropic response for each of the four NG-LAMP crystals, extending the equations of Section~\ref{sec:methods} from $I$ to $4I$ measurements and from one to four background rates---see Ref.~\cite[Sec.~II-A]{Hellfeld2019}.
Angular response functions $\eta(\theta, \phi)$ are computed at various photon energies using Geant4~\cite{Agostinelli2003, allison2006geant4, allison2016recent} models of the NG-LAMP detector system---see Figs.~\ref{fig:ng_render} and \ref{fig:ang_resp_2}.
The high-resolution simulated response functions were then scaled to match coarse experimental efficiency measurements.
Coordinate transforms were then applied to determine effective areas in terms of IMU-measured positions and orientations $\eta(\vec{r}_s, \vec{r}_i, \boldsymbol{q}_i$) from $\eta(\theta, \phi)$.

LiDAR-based SLAM~\cite{Hess2016} was performed to yield a 3D map of the environment and the time-dependent pose of the detection system.
Pose data was read out at approximately $4$~Hz, corresponding to an average distance between successive poses of ${\sim}10$~cm, depending on operator movement.
 \label{sec:nglamp}
\subsection{Measurements}
APSL was tested with two classes of experiments: source-search and source-separation (see the later Tables~\ref{tab:summary_search} and \ref{tab:summary_sep} for a list of runs).
Source-search measurements were performed in order to explore optimization and convergence behaviors, as well as spatial and intensity reconstruction accuracies, in scenarios with a non-trivial and unknown number of sources.
Source-separation measurements were performed in order to determine optimistic but representative spatial resolutions given realistic survey parameters, the type of detector system employed, and the selected source activities.

In the source-search measurements, up to four $8$~{\textmu}Ci \textsuperscript{137}{Cs} sources were placed throughout a laboratory at Lawrence Berkeley National Laboratory.
Four possible source locations on lab benches ${\sim}0.75$--$0.90$~m above the floor with separations of at least $2$~m were determined in advance in order to emulate four declared stations that could contain radiological sources.
While \textsuperscript{137}{Cs} was chosen for its availability and its clear $661.7$~keV photopeak, the inspection of declared stations might involve multiple gamma lines from U and/or Pu isotopes.

\begin{figure}[!htbp]
    \centering
    \includegraphics[width=\columnwidth]{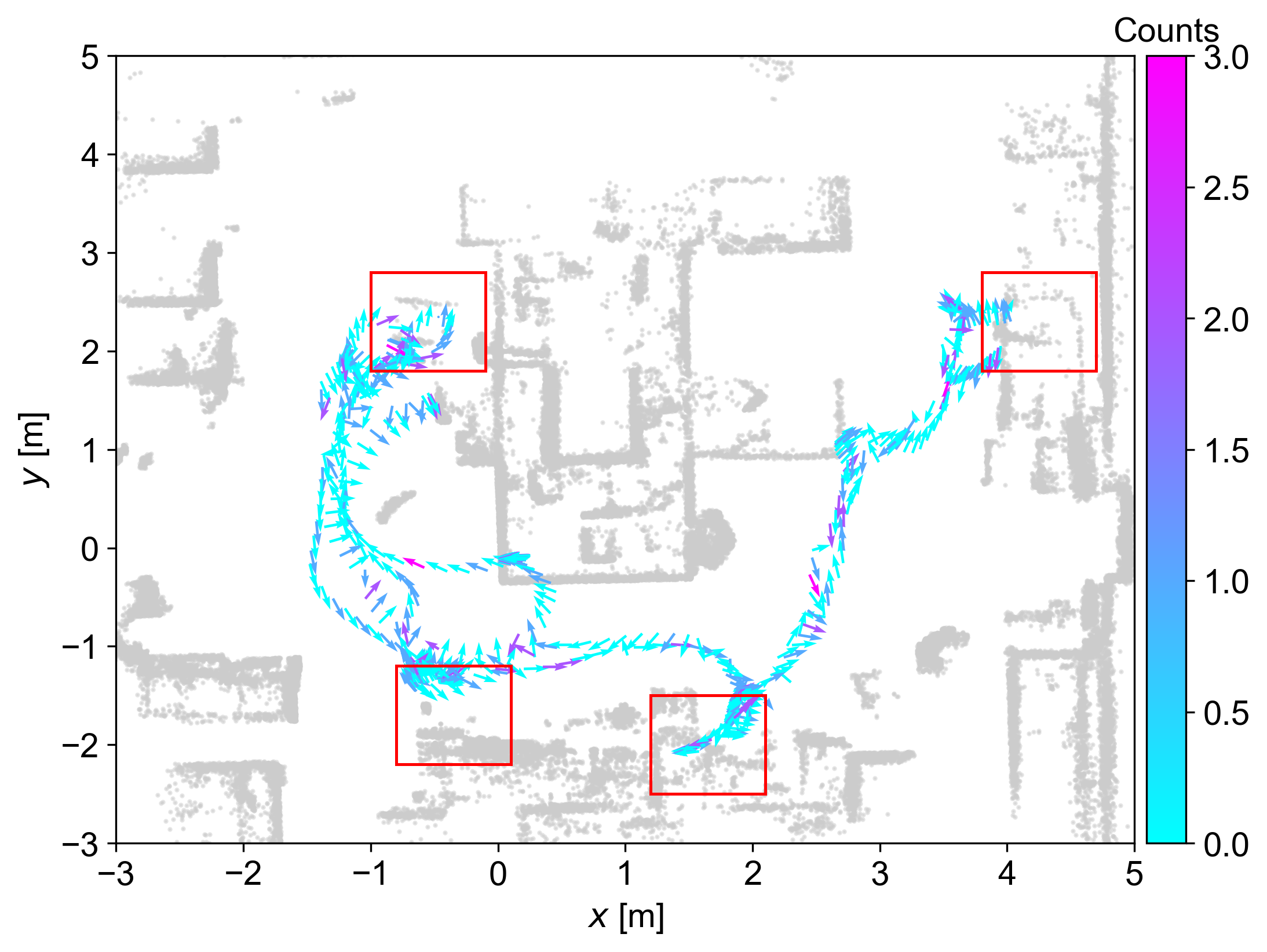}\\
    \includegraphics[width=0.95\columnwidth]{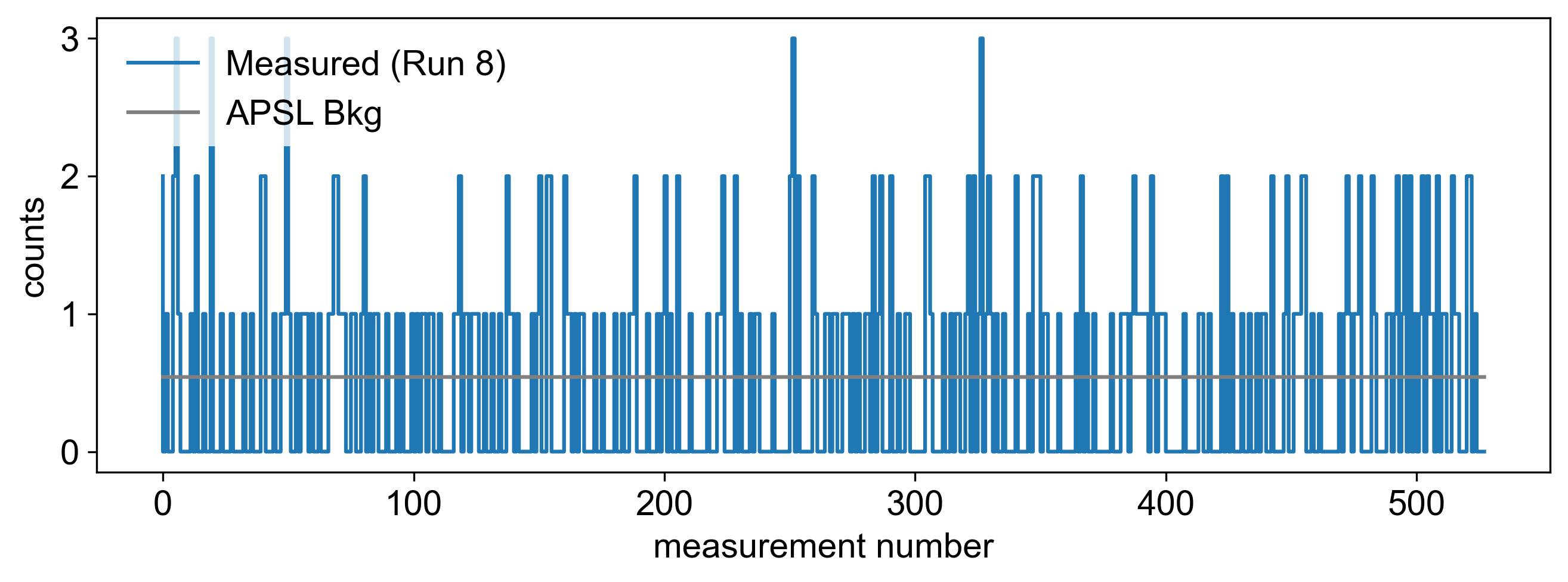}
    \caption{Top: top-down view of a LiDAR point cloud (grey points, downsampled $100\times$) and a typical search path (arrows) for the four declared areas (red rectangles) to be inspected.
    In this and subsequent figures, detector trajectories start and end near the origin.
    Arrows show the detector pose (position and orientation) every $0.25$~s and are colored by the counts in the region of interest (ROI) near $662$~keV (see Section~\ref{sec:recon}) and in the time window of the pose, summed over all four detectors.
    In this example, only background is present.
    Bottom: measured ROI counts, summed over all four detectors, as a function of detector pose number.}
    \label{fig:inspected_areas}
\end{figure}

A researcher acting as an inspector was then given a time limit of two to three minutes to inspect the four declared stations for the presence or absence of sources (and their positions and activities if present) using the NG-LAMP detector system.
The researcher was not informed which stations (if any) contained sources, and to further blind the search against visual data, the four possible source locations were obscured by cardboard boxes.
The researcher however had access to near-real-time (${\sim}2$~s delayed) count rate vs time data via the LAMP display to inform their search trajectory.
A typical search path and the set of search areas are shown in Fig.~\ref{fig:inspected_areas}.

In the source-separation measurements, two of the $8$~{\textmu}Ci \textsuperscript{137}{Cs} sources were placed at distances ranging from $0$ to $2.03$~m ($80$~inches) apart, in increments of $25.4$~cm ($10$~inches).
At each separation, the researcher was given one minute in which to freely move the detector throughout the scene (a `survey' pattern) to attempt a better characterization of the (visible) source positions.
In a similar set of runs at each separation, the researcher walked the detector past the sources four times (a `pass-by' pattern) at a constant height and an approximately fixed distance of $1$~m.
The survey patterns were designed to explore source separation in unconstrained scenarios where the researcher could 1) break measurement degeneracies by moving the detector across a wide range of $x$, $y$, and $z$ values; and 2) use visual and LAMP-provided count rate data to inform their search trajectory.
Conversely, the pass-by patterns were designed as a much more restrictive test of performance when feedback is disabled and measurements are highly degenerate. \label{sec:measurements}
\subsection{Reconstruction}
Measured counts $\boldsymbol{x}$ are computed by fitting each $^{137}$Cs $661.7$~keV photopeak (after gain stabilization) and summing counts within a region of interest (ROI) of $\pm\, 3\, \sigma$ around the peak centroid.
This standard deviation $\sigma$ is allowed to vary for each run and each detector in order to account for resolution differences among crystals---three of the NG-LAMP crystals have $\sigma \simeq 10$~keV, while one has $\sigma \simeq 14$~keV.
In the case of a poor fit result (e.g., if no $^{137}$Cs source is present), a fixed ROI of $\pm\, 30$~keV is used instead.
To reduce this reliance on sufficient counts in a particular photopeak, future applications might fit ROIs at multiple photopeak energies using calibration data ahead of time.

The minimization of Eq.~\ref{psloptimization} is then initialized with zero sources present and a suitable background estimate such as the median of the observed counts---see Algorithm~\ref{apslalgo}.
Starting points for new source parameters $\vec{r}_s, w_s$ are estimated using a gridded single-source version of point source localization (gPSL)~\cite[Sec.~III-A]{Hellfeld2019}, then passed to the NLopt~\cite{Johnson2008} COBYLA nonlinear optimization routine (parallelized via PYGMO~\cite{Biscani2019}) for the full APSL minimization.
The nonlinear optimizer is run until the optimal positions $\vec{r}_s$ are determined to within a relative tolerance of $10^{-3}$.
At each trial $\vec{r}_s$, the optimal $w_s$ is determined using 20~iterations of ML-EM.
If the BIC of the $(n+1)$-source model is greater than that of the $n$-source model, or if the $p$-value of the $n$-source model exceeds $0.05$, the $n$-source model is accepted and returned.

All reconstructions in this work were run offline, parallelized on a 12-thread 2.6~GHz Intel Core i7 processor.
With this hardware, APSL finds solutions in walltimes on the order of a minute, depending on the number of sources present, ML-EM iterations specified, and relative tolerances required.
The four-source run~7, for instance, required $38$~s for the APSL reconstruction, while the background-only run~8 terminated with a $p$-value of $0.425$ in $0.006$~s.

As two additional points of comparison for the final APSL reconstructions, we also compute 1) an ML-EM reconstruction~\cite[Sec.~II-B]{Hellfeld2019} with 200~iterations and a cubic voxel size of $20$~cm; and 2) in the source-search runs, the forward projection of the ground truth source locations and activities into count space, using the best estimates of source locations from the LiDAR point clouds and the mean source activity of $7.98$~{\textmu}Ci.
Ground truth positions were noted prior to the inspector's measurements, and their spatial coordinates were estimated in the point cloud of run~7 (Fig.~\ref{fig:run7}), which contained all four sources.
Coordinate frames of the remaining runs were then transformed to match the run~7 reference frame in order to use a constant set of ground truth locations.
Further discussion of this ground truth estimation procedure and associated uncertainties can be found in Section~\ref{sec:systematics}.
 \label{sec:recon}
\section{Results}\label{sec:results}

\subsection{APSL reconstructions}
Here, we evaluate the source-search and source-separation performance of APSL experimentally.
In the source-search measurements, performance metrics include whether APSL reconstructed the correct number of sources, and further, the errors in reconstructed source positions and activities from their respective ground truth values.
In the source-separation measurements, we again compare the number of sources, but then focus on the accuracy of the separation distance from its ground truth value instead of directly comparing each source position to its ground truth location.
Similarly, we compare the summed reconstructed activity against its ground truth value instead of comparing each source activity individually.

\begin{figure}[!htbp]
    \centering
    \includegraphics[width=0.95\columnwidth]{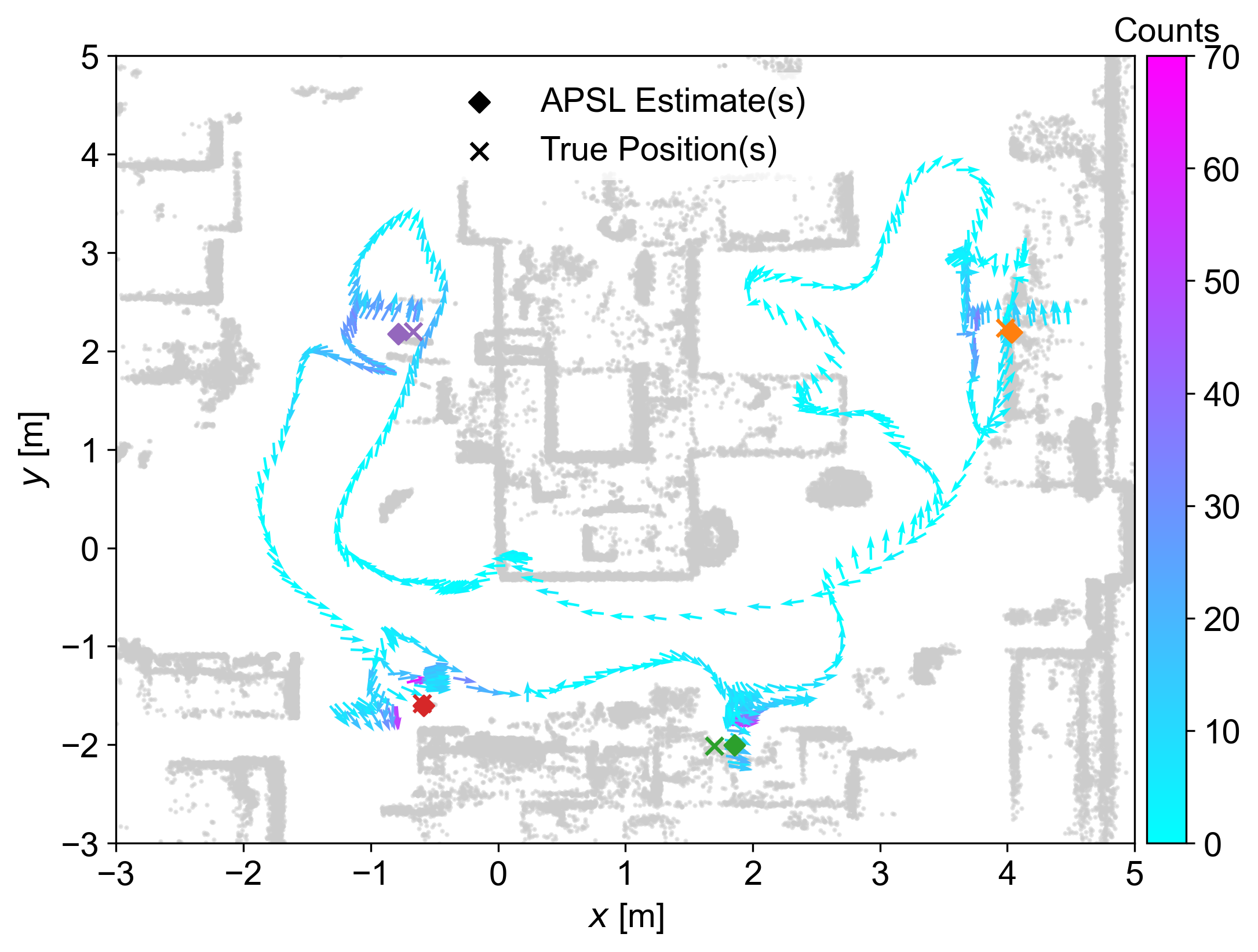}
    \includegraphics[width=0.95\columnwidth]{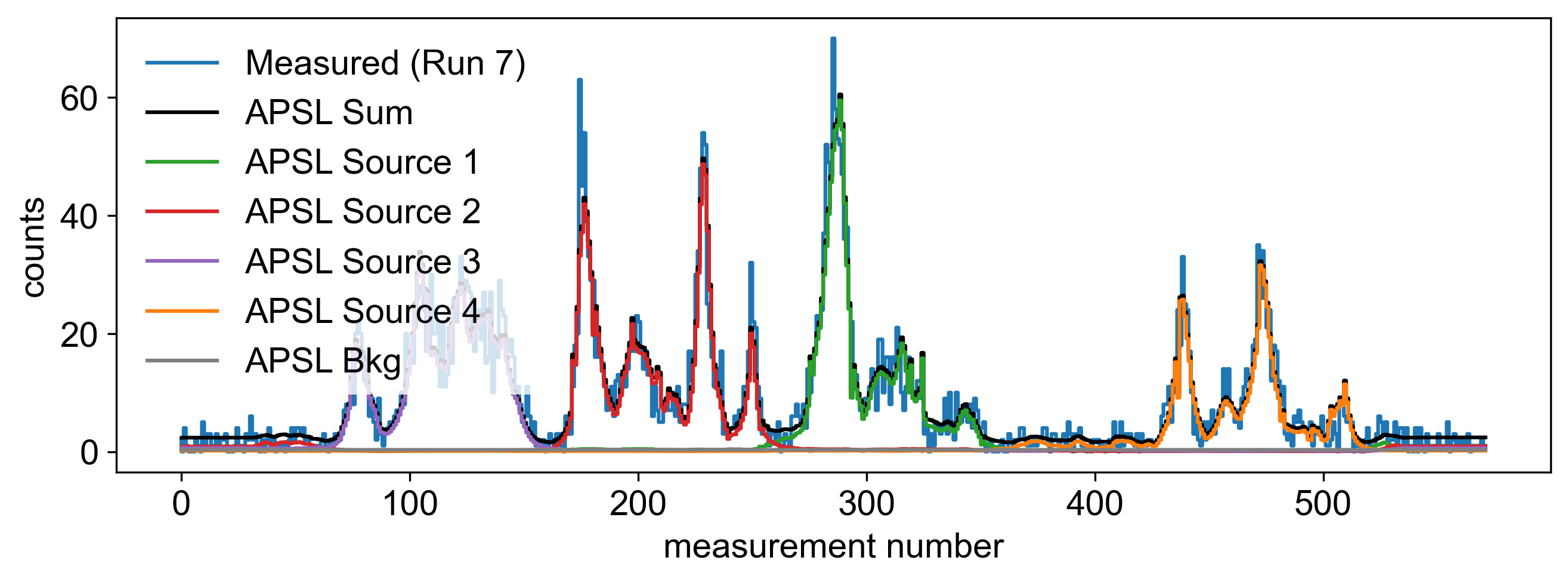}
    \includegraphics[width=0.95\columnwidth]{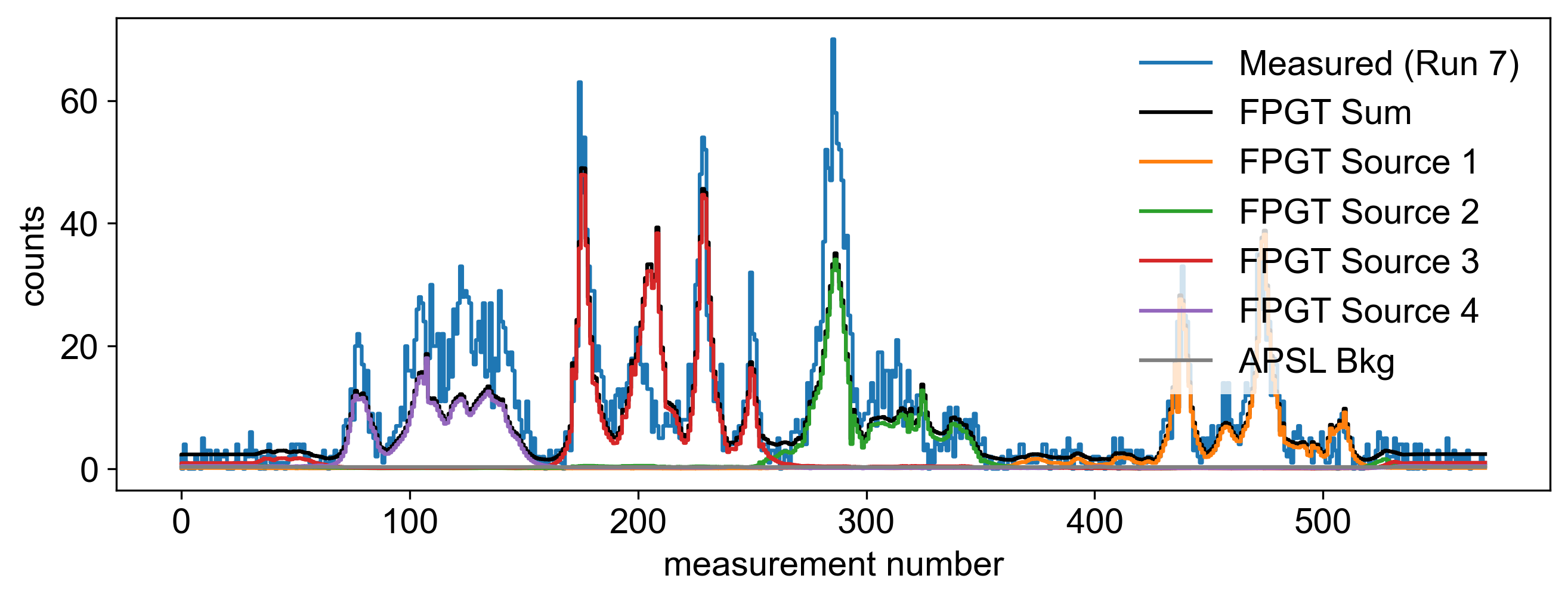}
    \caption{Top: detector trajectory, and reconstructed and ground truth source positions in run~7.
    Middle: reconstructed counts, summed over all four detectors. Bottom: forward projection of the ground truth (FPGT) source locations and activities.
    Here and in Fig.~\ref{fig:run3}, the APSL and FPGT sources are indexed differently but are plotted with matching colors.
    A 3D view of the scene around APSL source~\#3 near $(x,y) = (-1~\text{m},\,2~\text{m})$ is given in Fig.~\ref{fig:capture3D}.}
    \label{fig:run7}
\end{figure}

\begin{figure}[!htbp]
    \centering
    \includegraphics[width=\columnwidth]{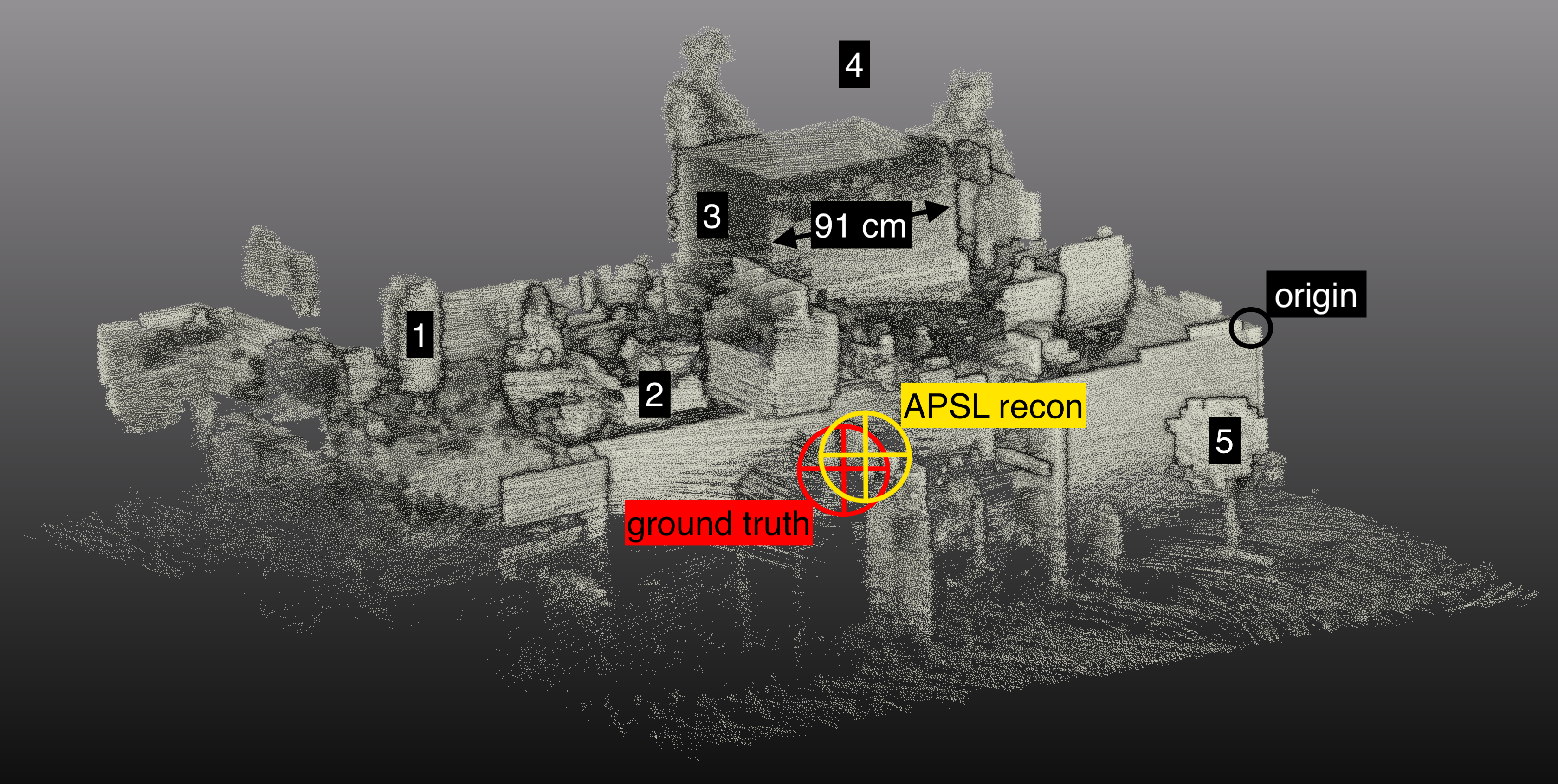}
    \caption{3D point cloud, ground truth source location (red crosshair), and reconstructed source location (yellow crosshair) of APSL source~\#3 near $(x,y) = (-1~\text{m},\,2~\text{m})$, from run~7 in Fig.~\ref{fig:run7}.
    The crosshair size is chosen only for visual clarity, and is much larger than the ${\sim}1$~cm spread in reconstructed positions over multiple random seeds.
    Several features in the point cloud are labelled for context: 1) a liquid nitrogen Dewar; 2) a cluttered laboratory bench; 3) a large concrete pillar, $91$~cm across; 4) two researchers standing in the background; and 5) a swivel chair.
    The approximate location of the coordinate origin in the source-search runs is also shown.
    The voxelized appearance of the chair and other features is the result of a $5$~cm moving voxel filter applied to reduce the number of noise points.}
    \label{fig:capture3D}
\end{figure}

Figs.~\ref{fig:run7} and \ref{fig:capture3D} show an experiment (run~7) in which APSL correctly determined that four \textsuperscript{137}{Cs} sources were present.
The errors in the spatial reconstruction were on average $16$~cm, compared to distances of closest approach of ${\sim}30$~cm, while the absolute errors in reconstructed activities were on average $0.9$~{\textmu}Ci or ${\sim}11\%$---see the later Table~\ref{tab:summary_search} for more detail.
The counts from each reconstructed source generally show close agreement with the measured counts, as do the counts computed from the forward projection of the ground truth source positions and activities.

\begin{figure}[!htbp]
    \centering
    \includegraphics[width=0.95\columnwidth]{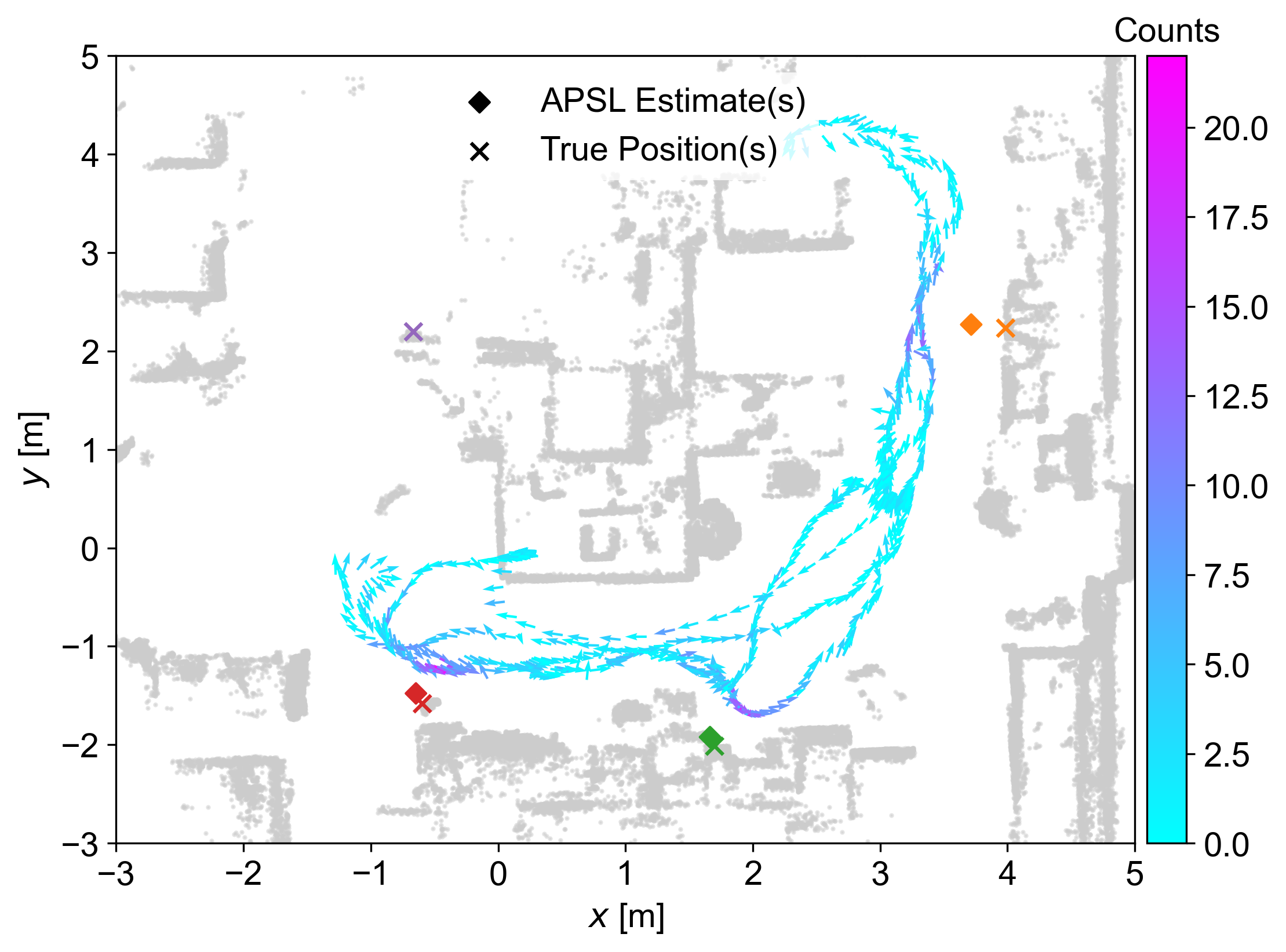}
    \includegraphics[width=0.95\columnwidth]{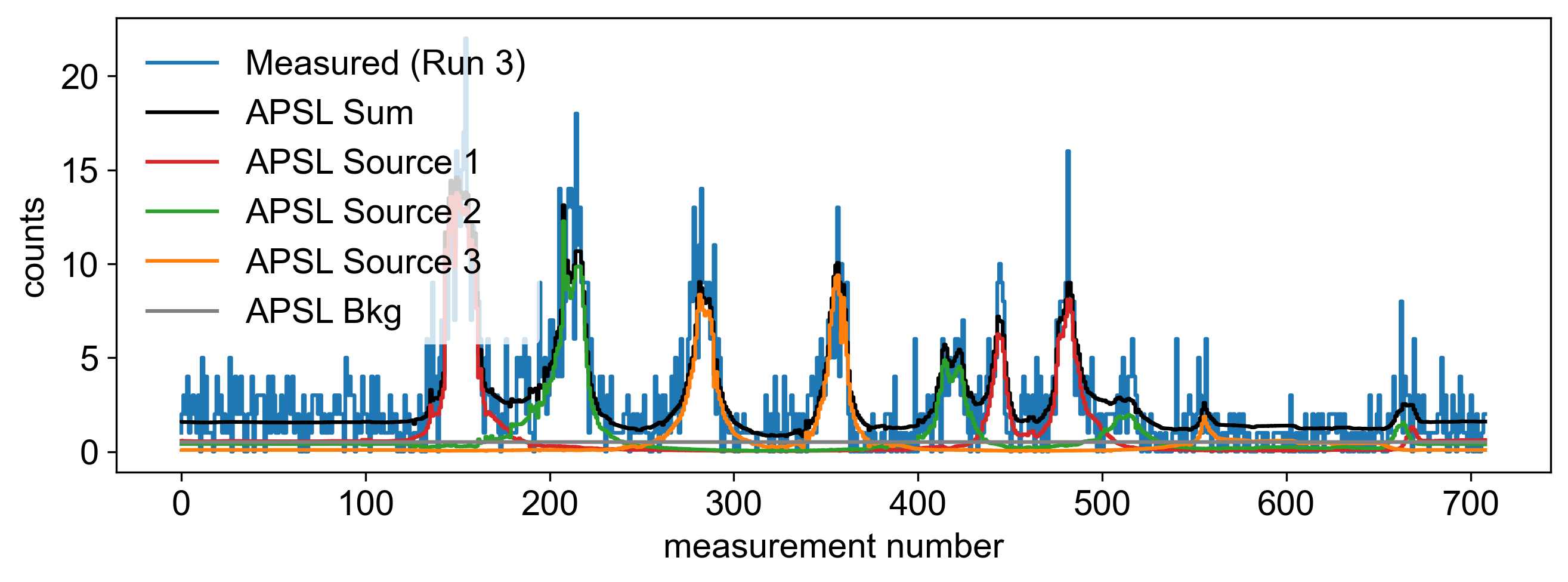}
    \includegraphics[width=0.95\columnwidth]{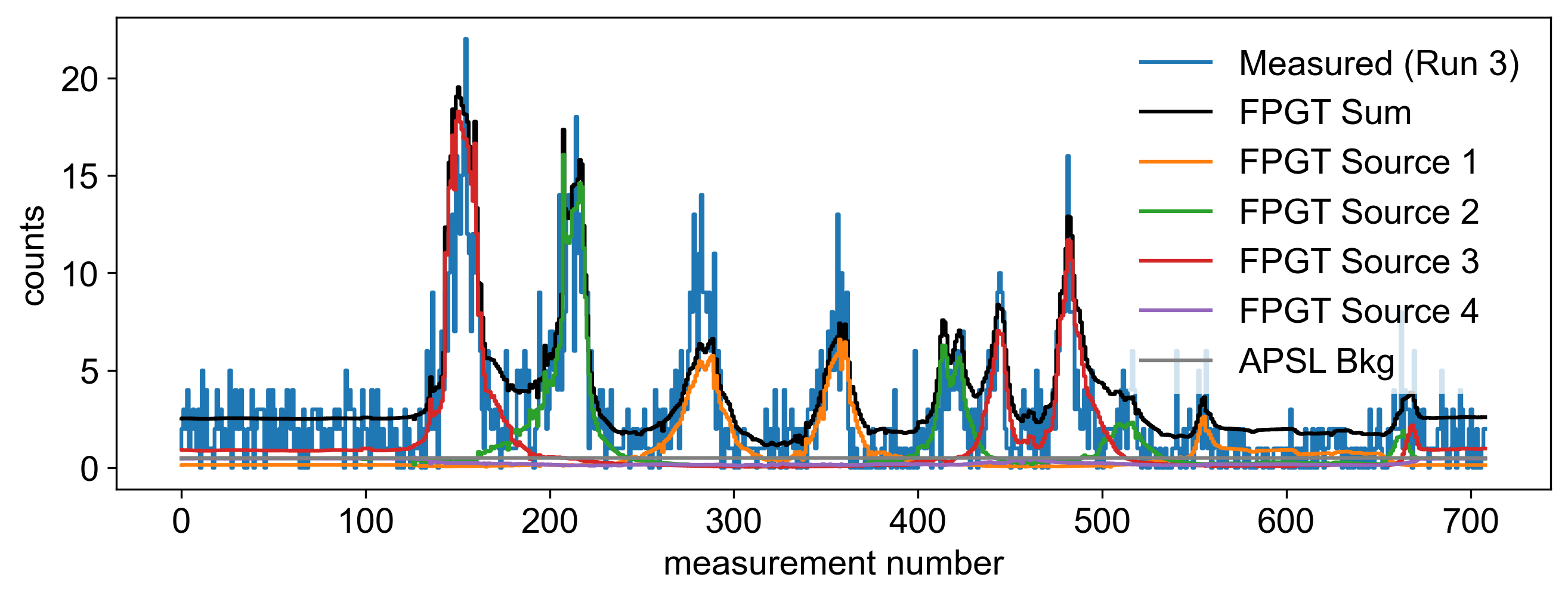}
    \caption{Top: detector trajectory, and reconstructed and ground truth source positions in run~3. Middle: reconstructed counts, summed over all four detectors. Bottom: forward projection of the ground truth (FPGT) source locations and activities.}
    \label{fig:run3}
\end{figure}

Fig.~\ref{fig:run3} shows another experiment (run~3) in which APSL reconstructed three out of the four true source locations, but failed to reconstruct the weakly-contributing fourth source location near the point $(-1~\text{m},\,2~\text{m})$, about $2.3$~m in closest approach from the detector trajectory.
The $xy$ position error of $27$~cm for the source near $(4~\text{m},\,2~\text{m})$ accounts for almost the entire $xyz$ position error on the source, and is the largest observed $xy$ position error across the set of source-search runs, possibly due to the degenerate detector trajectory near the source.

The earlier Fig.~\ref{fig:inspected_areas} shows a background-only experiment (run~8) where APSL correctly reconstructs a scenario with zero sources present.
As the ML-EM steps within APSL preserve the number of counts, the reconstructed constant background rate is simply the average rate of the data.

\begin{figure}[!htbp]
    \centering
    \includegraphics[width=\columnwidth]{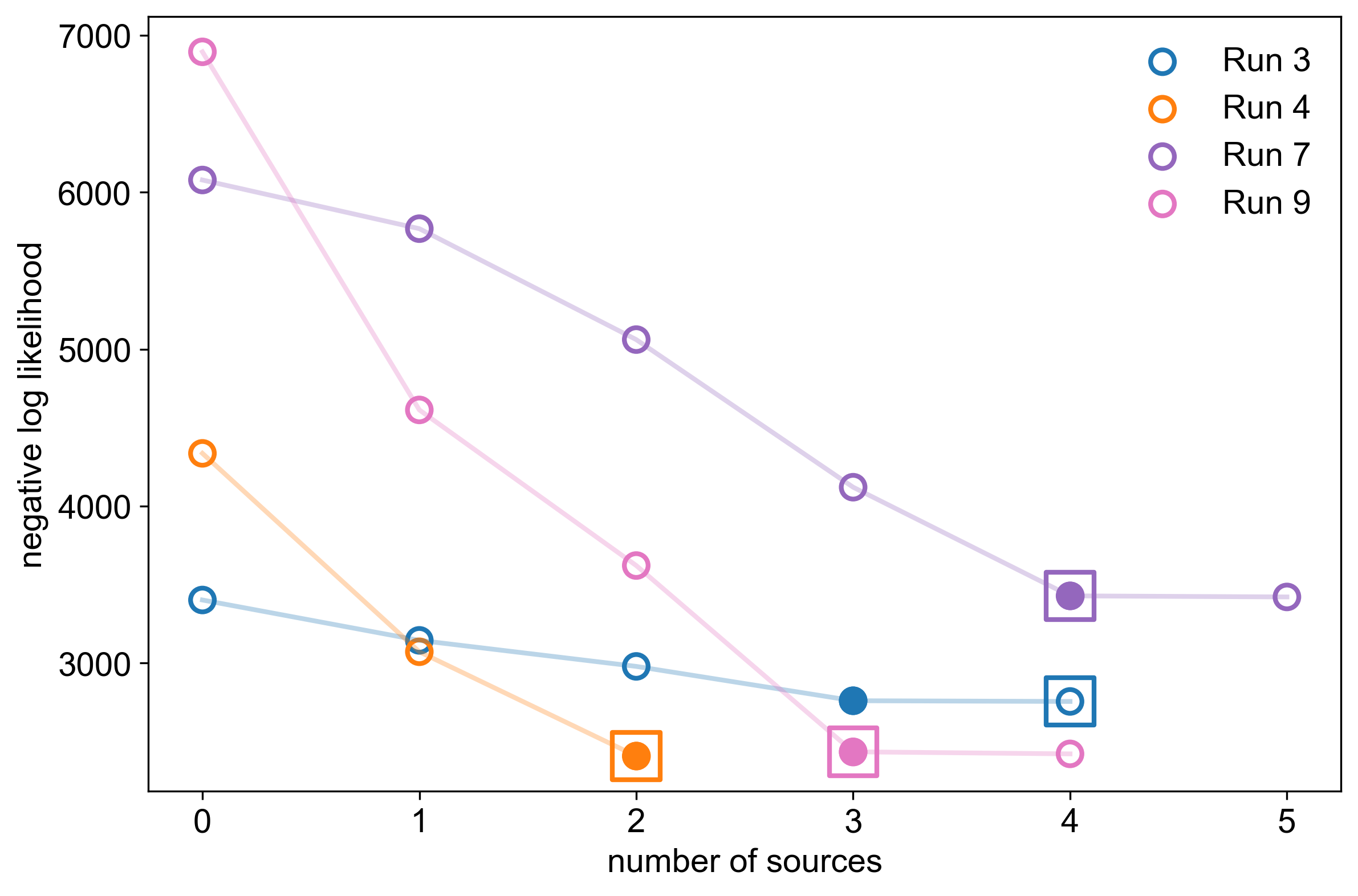}
    \caption{Evolution of the negative log-likelihood for runs~3, 4, 7, and 9. Open circles denote rejected models, filled circles denote final accepted models, and squares denote the true number of sources. Lines between points are drawn only to guide the eye.}
    \label{fig:ll_evolution_1}
\end{figure}

Fig.~\ref{fig:ll_evolution_1} shows the evolution of the optimum negative log-likelihood $\nll$ vs the number of modeled sources for four source-search experiments.
In most source-search reconstructions (runs~3, 5--7, 9--11), $\nll$ falls roughly linearly with the number of sources, then saturates abruptly when adding one more source than necessary only marginally improves the model and fails to improve the BIC.
We note again that run~3 terminates one source too early as a result of its search path not approaching one of the sources.
Runs~4, 8, and 10, conversely, stop based on a sufficiently large $p$-value, rather than the BIC criterion.
We also note some sensitivity to the choice of random seed used in selecting initial parameter values via gPSL (see Section~\ref{sec:recon}).
In run~5, for instance, approximately $18\%$ of reconstructions with different random seeds split the source near the point $(4~\text{m},\,2~\text{m})$ into one strong and one weak source.
For concreteness, results presented here are the most common reconstruction modes.

Across the nine source-search runs (Table~\ref{tab:summary_search}), eight had sources present.
In eight of the nine runs, APSL reconstructed the correct number of sources, with run~3 (Fig.~\ref{fig:run3}) as the sole exception.
In all eight source-present runs, every reconstructed source position was determined to within $46$~cm of a ground-truth source position.
In seven of eight source-present runs, every reconstructed position was within the distance of closest approach of the closest ground-truth position.
This position accuracy relative to the distance of closest approach ranged from $29\%$ to $146\%$, with an average of $60\%$.
The average position and (unsigned) activity errors across the eight source-present runs were $18$~cm and $1.6$~{\textmu}Ci ($20\%$), respectively.

We also note that the proper accounting of individual anisotropic detector response functions (e.g., Fig.~\ref{fig:ang_resp_2}) is important for the robustness of the APSL reconstructions.
For comparison, we also considered the union of detectors by summing the response~$\eta$ and ROI counts~$\boldsymbol{x}$ across the four NG-LAMP crystals.
In the source-search runs, only six of nine APSL reconstructions with this unified model (and the same random seed) produced the correct number of sources, compared to eight of nine with the full model.
We also ran APSL with an isotropic and unified response, where we replaced the unified $\eta(\theta, \phi)$ with its average over all $\theta, \phi$.
With this yet-simpler model, only five of nine reconstructions (again with the same seed) returned the correct number of sources.
Sweeping over multiple random seeds, we find that in run~5, for instance, with two true sources, the full model reconstructed $\{2, 3\}$ sources $\{82, 18\}\%$ of the time.
The unified model by contrast reconstructed $\{1, 2, 3, 4\}$ sources in $\{40, 40, 16, 4\}\%$ of trials, while the unified+isotropic model performed worse still with rates of $\{44, 28, 20, 8\}\%$.
All other results presented in this work therefore use the non-unified, anisotropic responses~$\eta(\theta, \phi)$.

In the nine source-separation experiments conducted using a survey pattern (see Section~\ref{sec:measurements}), APSL reconstructed two discrete sources in six of the eight runs that had a non-zero true separation.
In these six survey runs (19, 21, 23, 25, 27, 29), the reconstructed separations were within $15$~cm of the true separation.
The minimum true separation for which this level of accuracy was achieved was $76$~cm (run~19), with an average and standard deviation reconstructed separation over 10 random seeds of $82 \pm 2$~cm.
We therefore interpret $76$~cm as the approximate spatial resolution for $8$~{\textmu}Ci activities using the NG-LAMP detector system.
This resolution will differ for other detector systems, and will likely improve for stronger sources.

We also find that high pose variability and especially high signal-to-noise are crucial for accurately reconstructing multiple spatially-close sources, as APSL was unable to reconstruct two discrete sources in the nine pass-by runs---see Section~\ref{sec:future_work} for further discussion.
Instead, APSL typically reconstructed a single source near the midpoint of the two ground-truth source locations, with a source activity roughly double the individual true source activities.

\subsection{Comparison to ML-EM reconstructions}
\begin{figure}[!htbp]
    \centering
    \includegraphics[width=\columnwidth]{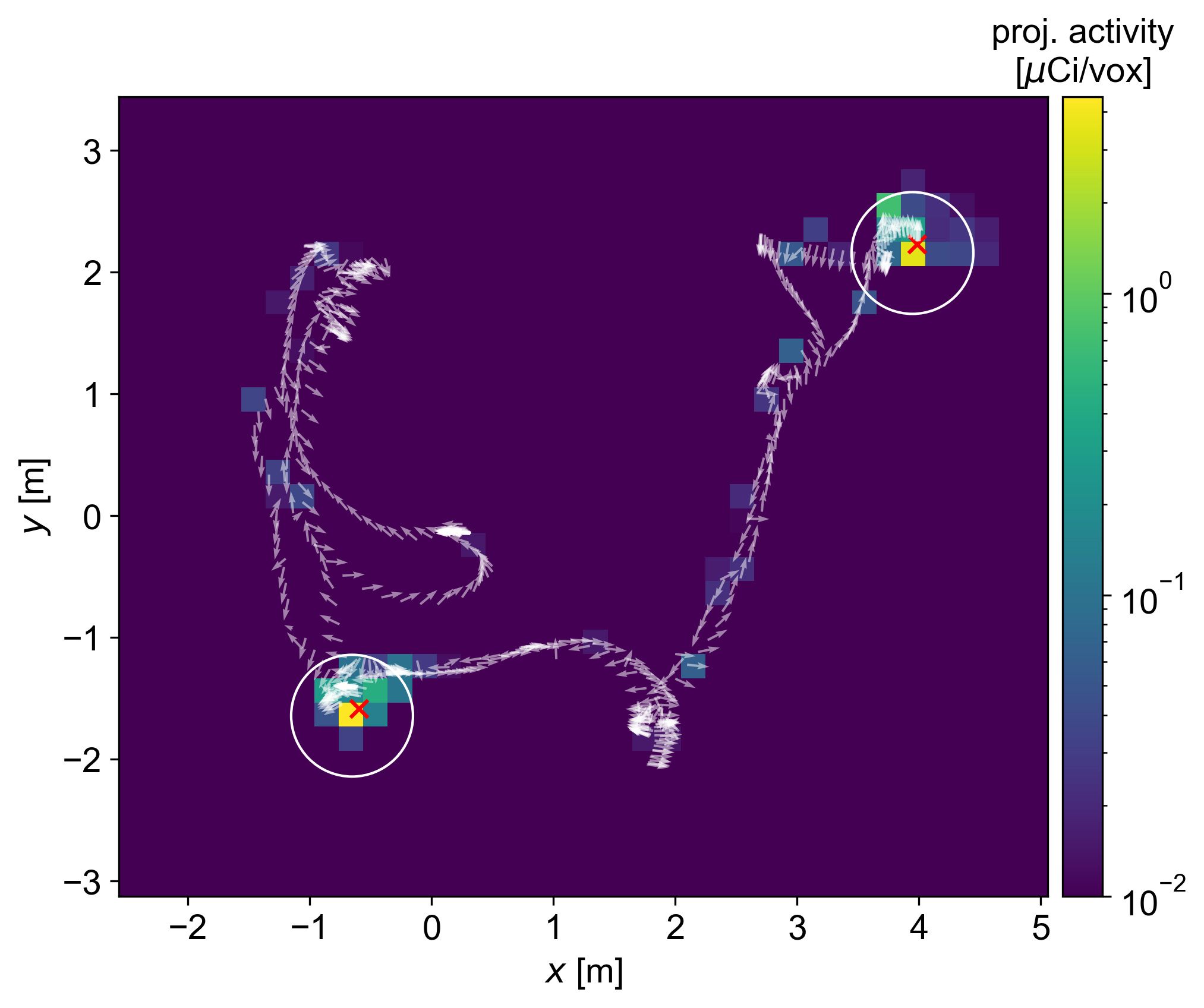}\\
    \includegraphics[width=\columnwidth]{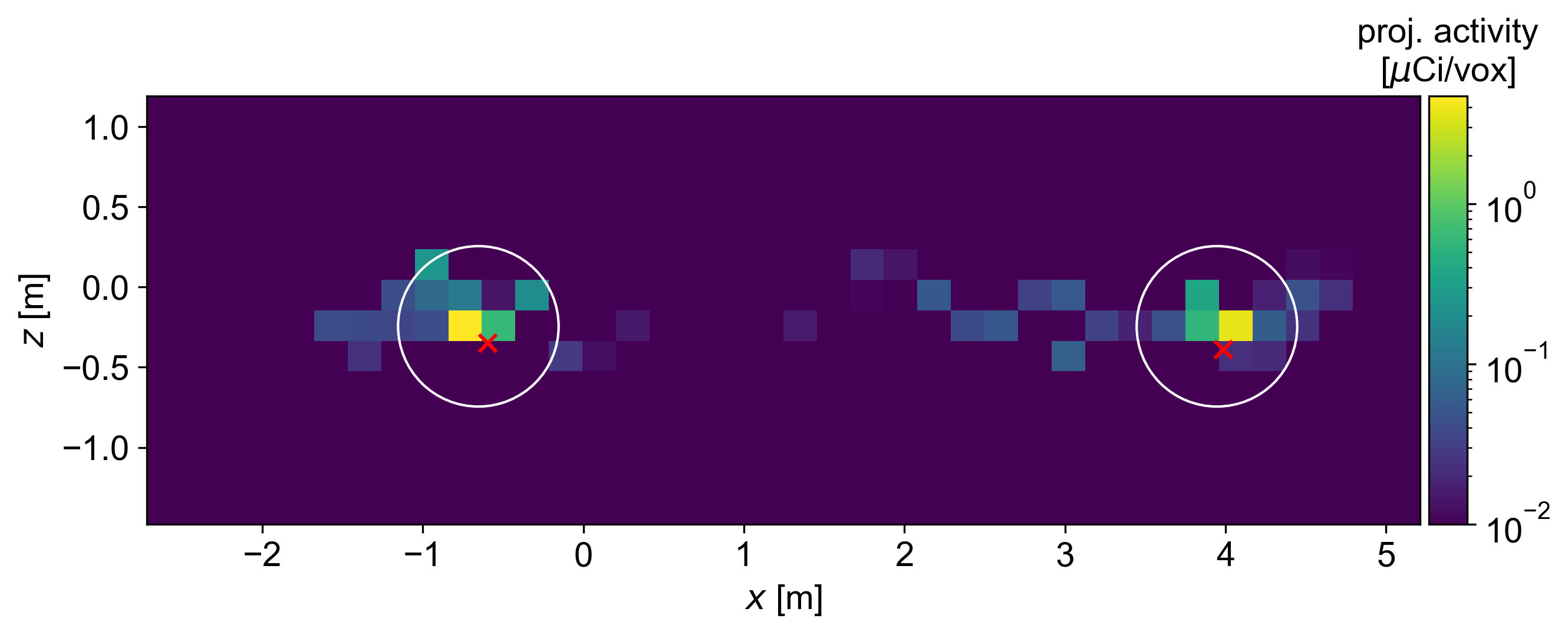}
    \caption{ML-EM reconstructions for run~5, using 200 iterations and a 20~cm voxel size, shown as projections along the $z$-axis (top) and $y$-axis (bottom).
    Small projection activities below $10^{-2}$~{\textmu}Ci are thresholded to $10^{-2}$~{\textmu}Ci for visual clarity.
    Detector poses (top plot only) are shown as semi-transparent white arrows, and true source locations are shown as red $\times$ markers.
    White circles are drawn at the $50$~cm radius used to define the activity of each ML-EM-reconstructed source.}
    \label{fig:run5_mlem}
\end{figure}

\begin{figure}[!htbp]
    \centering
    \includegraphics[width=\columnwidth]{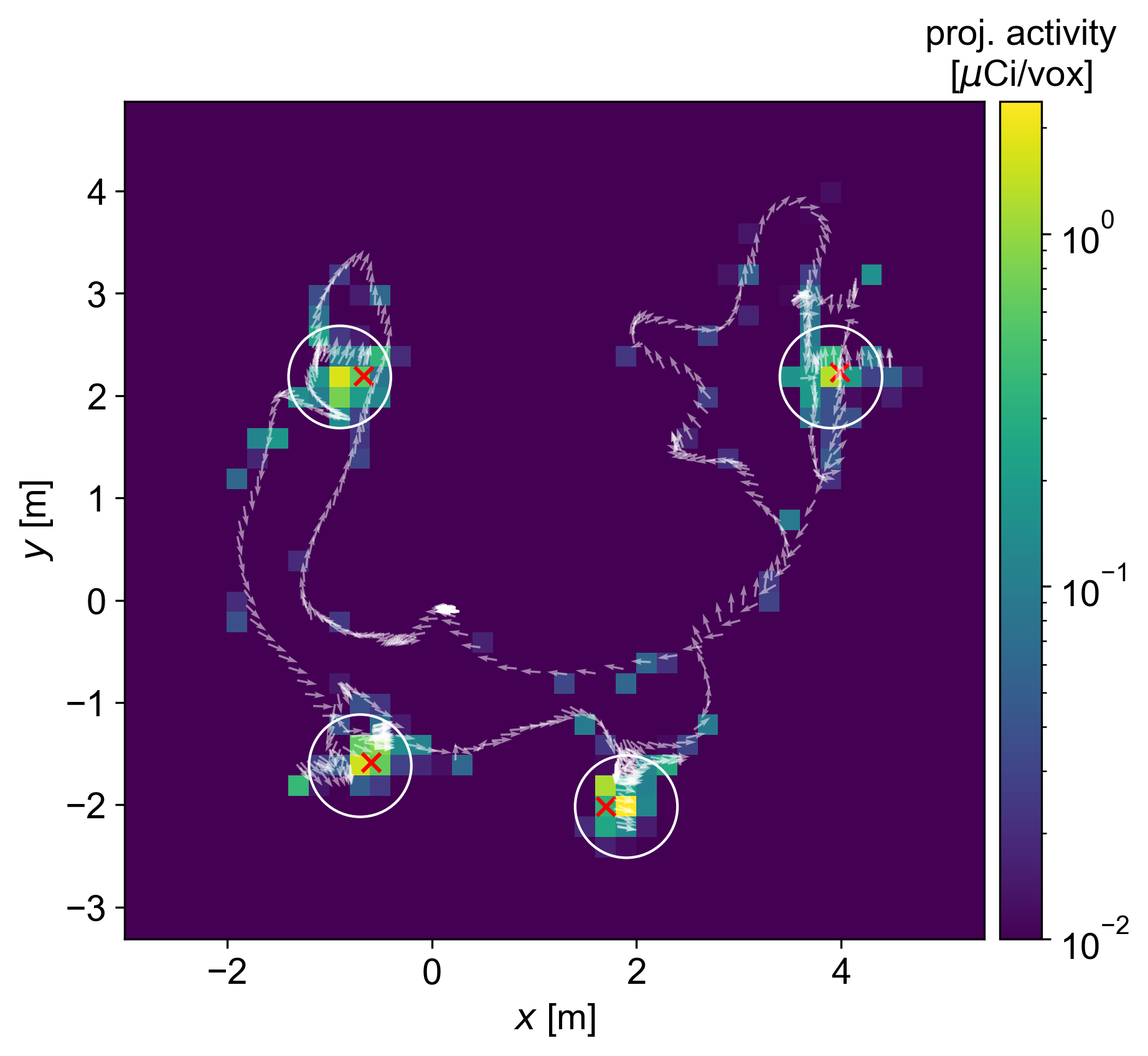}\\
    \includegraphics[width=\columnwidth]{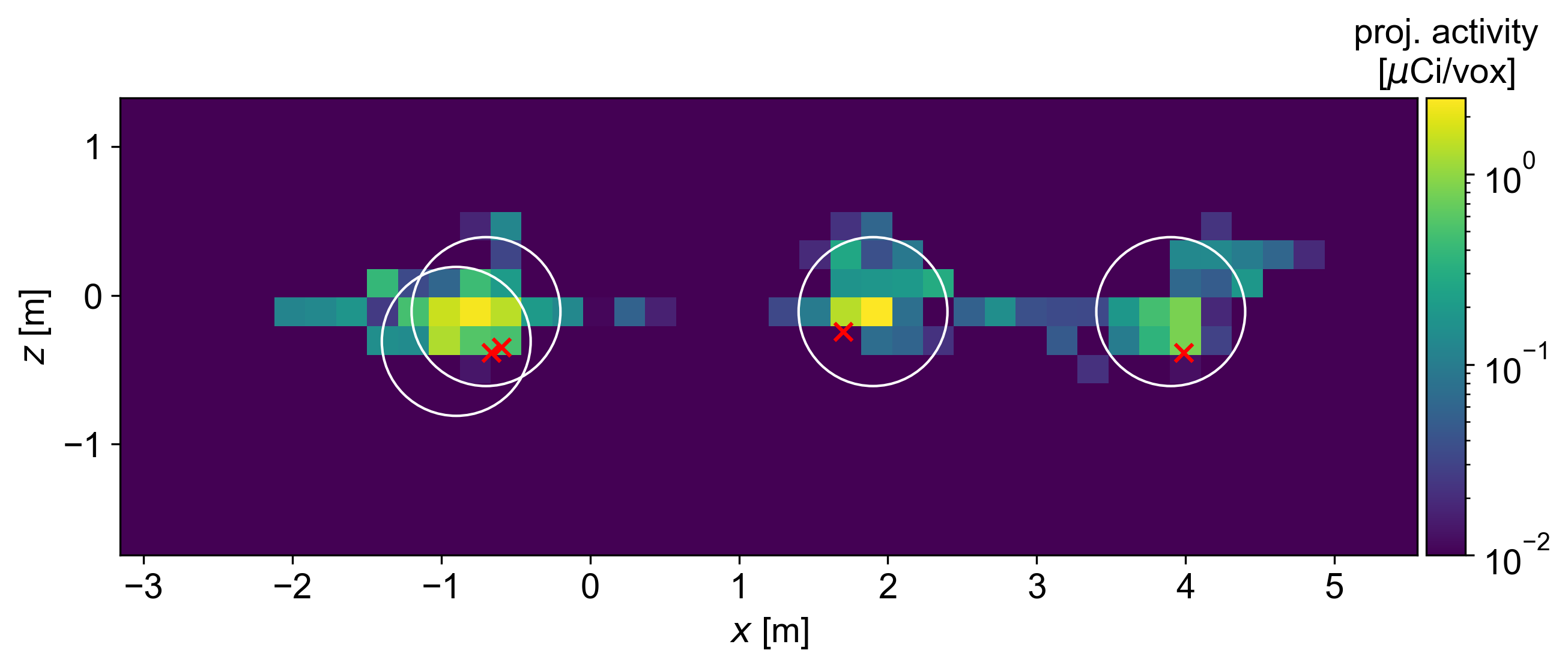}
    \caption{ML-EM reconstructions for run~7, using the same parameters as Fig.~\ref{fig:run5_mlem}.}
    \label{fig:run7_mlem}
\end{figure}

Performing ML-EM reconstructions of the source-search runs offers some additional insight, but requires some choice in interpretation.
As shown for runs~5 and 7 in Figs.~\ref{fig:run5_mlem} and \ref{fig:run7_mlem}, the ML-EM reconstructions do successfully produce localized regions of activity near the ground truth source positions.
However, the ML-EM activities are spread across multiple voxels over distances of ${\sim}50$~cm or more.
To facilitate comparisons of ML-EM position and activity accuracy against ground truth values and APSL results, we define the position of an ML-EM-reconstructed source to be the voxel center of the highest-activity voxel within $50$~cm of the voxel center of each ground truth source.
This definition makes use of ground truth position information and so is unsuitable for true search scenarios; however, it allows us to quantify accuracies in this work without resorting to peak detection or fitting methods.
Using this voxel center, we then sum the activity in voxels with centers ${\leq}50$~cm away to define the activity of an ML-EM-reconstructed source.

With these definitions, we find that the ML-EM reconstructions of the source-search runs occasionally produce accurate results, but more often give larger position errors and substantially underestimated activities compared to APSL.
Fig.~\ref{fig:run5_mlem} (run~5) shows one of the more accurate ML-EM reconstructions, with average position errors and activity errors of $15$~cm and $1.6$~{\textmu}Ci ($20\%$), compared to APSL errors of $12$~cm and $0.8$~{\textmu}Ci ($10\%$).
Fig.~\ref{fig:run7_mlem} (run~7) shows a more typical ML-EM reconstruction, with position errors of $26$~cm and activity errors of $3.1$~{\textmu}Ci ($39\%$), compared to the aforementioned run~7 APSL errors of $16$~cm and $0.9$~{\textmu}Ci ($11\%$).
Across all eight source search runs in which at least one source was present, the ML-EM reconstructions produce average position and (unsigned) activity errors of $26$~cm and $3.6$~{\textmu}Ci ($46\%$); as discussed above, the corresponding average values for APSL are $18$~cm and $1.6$~{\textmu}Ci ($20\%$).
The mean (standard deviation) wall time required for the APSL and ML-EM reconstructions are similar at $22\,(9)$~s and $21\,(4)$~s, respectively.

In the source-separation runs, the relatively small distance between sources in most runs produces highly blurred ML-EM images, making the application of the aforementioned position and activity definitions difficult.
Rather than attempting to use these definitions, we provide some more general results.
First, as with the source-search ML-EM reconstructions in Figs.~\ref{fig:run5_mlem} and \ref{fig:run7_mlem}, reconstructed activity is spread across multiple voxels.
This spread is especially large for the pass-by runs, where nearly all the reconstructed activity is distributed along or next to several meters of the detector trajectory.
In the survey runs, the ML-EM reconstructions produce activity distributions consistent with two distinct sources starting around separations of 76~cm---similar to APSL---but distinguishing closely spaced ML-EM-reconstructed sources without prior knowledge of the ground truth is in general difficult.
Interestingly, the total-image ML-EM activities in the survey runs are closer on average to the true value of $16.6$~{\textmu}Ci than are the APSL activities (see Table~\ref{tab:summary_sep}).
By contrast, the total-image ML-EM activities in the pass-by runs are significantly underestimated, with no run surpassing a total of $6.6$~{\textmu}Ci.

Results for both APSL and ML-EM reconstructions are tabulated in greater detail in Tables~\ref{tab:summary_search} and \ref{tab:summary_sep}.
The diagnostic runs~0--2 and pass-by runs (even numbers from 12--28) are not included in the tables.
Values are given for a single representative reconstruction (i.e., a single random seed) of each run.
Runtimes of the measurements, APSL reconstructions, and ML-EM reconstructions are given for comparison; the timing of the latter includes both the time required to compute the system response matrix as well as to perform the 200 ML-EM iterations.
The ROI counts $\sum_i x_i$ are the total counts within $\pm\, 3\, \sigma$ of the $661.7$~keV photopeak fits during the entire measurement time, while the background column $b$ gives the APSL-reconstructed background rate in the ROI in counts per second (rather than per pose).
The maximum signal-to-noise ratio is defined as
\begin{align}
    \text{SNR}_\text{max} = \text{max}_i \{ \left(x_i - b t_i\right) / \sqrt{x_i} \}
\end{align}
over all poses $i$, with $x_i$ and $b$ summed over the four detectors.
APSL position errors $r_\text{err}$ are computed based on the closest ground-truth position.
APSL activity errors $w_\text{err}$ are computed similarly as ground truth minus reconstructed, using the average source intensity (corrected for decay time) as the ground truth estimate for all sources.
ML-EM position and activity errors $r_\text{err}$ and $w_\text{err}$ are computed based on the hottest voxel position and total activity within a $50$~cm radius as discussed above.
The total activity $w_\text{tot}$ in Table~\ref{tab:summary_sep} is the sum of all APSL-reconstructed activities, the ground truth value of which is $16.6$~{\textmu}Ci.
All activities and errors are given in terms of $^{137}$Cs nuclear disintegrations, $85.1\%$ of which produce a $661.7$~keV photon~\cite{Browne2007}.

\begin{table*}[t]
    \centering
    \caption{Summary of source-search runs}
    \setlength{\tabcolsep}{0.4em}
    \begin{tabular}{c|cccccccllll}
        run & meas & ROI & $b$ & SNR$_\text{max}$ & recon/true & recon time & recon time & $r_\text{err}$ & $r_\text{err}$ & $w_\text{err}$ & $w_\text{err}$ \\
        \# & time [s] & counts & [cps] & --- & srcs (APSL) & (APSL) [s] & (ML-EM) [s] & (APSL) [cm] & (ML-EM) [cm] & (APSL) [{\textmu}Ci] & (ML-EM) [{\textmu}Ci] \\\hline
3 & 177 & 1922 & 2.0 & 4.6 & 3/4 & 27.6 & 18.3 & 19, 32, 27 & 51, 43, 21, 46 & 3.5, 1.9, 3.6 & 6.2, 6.2, 4.8, 8.0 \\
4 & 154 & 2544 & 1.8 & 6.2 & 2/2 & 23.1 & 27.2 & 11, 20 & 18, 8 & 1.1, 1.2 & 4.5, 3.1 \\
5 & 139 & 4044 & 1.3 & 9.2 & 2/2 & 18.8 & 14.8 & 12, 12 & 16, 13 & 1.1, -0.4 & 2.1, 1.1 \\
6 & 136 & 3530 & 2.1 & 11.4 & 1/1 & 10.7 & 17.0 & 16 & 21 & 2.0 & -0.2 \\
7 & 143 & 5175 & 1.2 & 8.3 & 4/4 & 37.9 & 23.4 & 23, 19, 14, 7 & 29, 24, 26, 25 & 0.2, 0.6, -1.9, -0.7 & 4.7, 2.3, 3.2, 2.3 \\
8 & 132 & 286 & 2.2 & 1.4 & 0/0 & 0.0 & 15.5 & --- & --- & --- & --- \\
9 & 134 & 4047 & 0.4 & 9.7 & 3/3 & 28.2 & 17.8 & 16, 46, 16 & 46, 22, 16 & -2.9, 1.6, 1.6 & 3.6, 4.2, -0.8 \\
10 & 141 & 3832 & 1.1 & 8.7 & 3/3 & 18.3 & 26.0 & 10, 14, 17 & 22, 9, 36 & -2.1, -1.1, 1.8 & 4.9, 0.7, 4.2 \\
11 & 132 & 1646 & 1.3 & 5.3 & 1/1 & 8.2 & 21.8 & 13 & 31 & 0.4 & 5.1 \\

    \end{tabular}
    \label{tab:summary_search}
\end{table*}

\begin{table*}[t]
    \centering
    \caption{Summary of source-separation runs}
    \setlength{\tabcolsep}{0.5em}
    \begin{tabular}{c|ccccccccccc}
        run & meas & ROI & $b$ & SNR$_\text{max}$ & recon/true & recon time & recon time & true sep & recon sep & $w_\text{tot}$ & $w_\text{tot}$ \\
        \# & time [s] & counts & [cps] & --- & srcs (APSL) & (APSL) [s] & (ML-EM) [s] & [cm] & (APSL) [cm] & (APSL) [{\textmu}Ci] & (ML-EM) [{\textmu}Ci]\\\hline
13 & 66 & 3280 & 0.8 & 9.1 & 1/2 & 7.8 & 5.8 & 0 & --- & 22.9 & 16.8 \\
15 & 60 & 2743 & 0.8 & 7.3 & 1/2 & 7.4 & 5.1 & 25 & --- & 21.5 & 16.3 \\
17 & 64 & 2275 & 0.7 & 5.3 & 1/2 & 7.6 & 5.0 & 51 & --- & 30.5 & 12.9 \\
19 & 71 & 2791 & 2.4 & 6.4 & 2/2 & 8.2 & 6.2 & 76 & 80 & 20.5 & 15.2 \\
21 & 71 & 2583 & 1.5 & 6.3 & 2/2 & 8.6 & 5.9 & 102 & 115 & 23.0 & 14.9 \\
23 & 62 & 2110 & 0.7 & 5.6 & 2/2 & 13.8 & 5.2 & 127 & 123 & 25.0 & 13.6 \\
25 & 62 & 1934 & 1.5 & 5.5 & 2/2 & 25.1 & 5.0 & 152 & 157 & 19.2 & 11.3 \\
27 & 68 & 1819 & 0.0 & 5.7 & 2/2 & 9.3 & 5.7 & 178 & 192 & 24.7 & 13.8 \\
29 & 66 & 2118 & 1.6 & 7.0 & 2/2 & 18.1 & 5.7 & 203 & 218 & 18.7 & 11.4 \\

    \end{tabular}
    \label{tab:summary_sep}
\end{table*}

\section{Discussion}\label{sec:discussion}

\subsection{APSL vs ML-EM}
As shown in Section~\ref{sec:results}, APSL outperforms ML-EM in position and especially activity reconstruction---the average position and (unsigned) activity errors were $26$~cm and $3.6$~{\textmu}Ci ($46\%$) for ML-EM, compared to $18$~cm and $1.6$~{\textmu}Ci ($20\%$) for APSL.
Here we expand on these results with some additional discussion.

We note that the large activity underprediction from ML-EM is not primarily a result of using an arbitrary radius of $0.50$~m for activity summing---increasing this radius to $1$~m, for instance, reduces the average unsigned activity error only slightly to $3.0$~{\textmu}Ci ($38\%$).
Similarly, adjusting the number of iterations in the ML-EM reconstruction to $500$, giving the algorithm more time to converge, produces an activity error of $3.5$~{\textmu}Ci ($43\%$).

Instead, these underpredictions are inherent to the voxellized, non-sparse model: the ML-EM reconstruction can assign activity to more sources than APSL, and many of these voxels will lie on or near the detector trajectory (see Fig.~\ref{fig:run7_mlem} in particular).
As a result of their smaller distance in the $1/r^2$ factor of Eq.~\ref{eq:sysresp}, these nearby voxels will require substantially lower reconstructed activity $w$ to account for the observed signal $\boldsymbol{x}$.

These ML-EM results highlight the superior applicability of APSL to truly sparse point source scenarios: with APSL, one need not define additional parameters such as the radius in which to sum voxel activities, nor worry about how to define and localize hotspots, nor deal with issues such as voxel size limitations on spatial accuracy.
APSL also offers a statistically-founded, unambiguous stopping criterion based on the Bayesian Information Criterion or $p$-value, rather than a typically arbitrary number of iterations in ML-EM.
Visualizing ML-EM reconstructions of sparse scenarios for human interpretation also tends to be more difficult and arbitrary than visualizing APSL reconstructions.
For instance, plotting ML-EM activities projected along the $z$-axis on a logarithmic color scale (e.g.~Figs.~\ref{fig:run5_mlem} and \ref{fig:run7_mlem}) helps visualize the distribution of the total activity across the space, but tends to visually blur activity hotspots compared to a single-voxel-width slice plotted on a linear color scale.

\subsection{Systematic uncertainties}\label{sec:systematics}
The reconstruction results and their comparison to ground truth values suffer from three main sources of systematic uncertainty: the accuracy of LiDAR point cloud alignments across the set of runs, the accuracy of ground truth source location estimates from the point clouds, and the far-field assumption in the detector response calculations.

Each point cloud has a different laboratory coordinate frame, depending on slight misalignments between the poses of the IMU/LiDAR at the start of each measurement.
In order to use a constant coordinate frame for the ground truth source locations, the source-search point clouds are aligned or co-registered with the point cloud of run~7 (which contained all four sources) using random sample consensus (RANSAC)~\cite{Fischler1981} followed by iterative closest point (ICP)~\cite{Besl1992} algorithms in Open3D~\cite{Zhou2018} and CloudCompare~\cite{CloudCompare}.
Imperfections in the co-registration transform and thus reconstructed positions contribute ${\lesssim}5$~cm to the position errors $r_\text{err}$ in Table~\ref{tab:summary_search}, but do not affect the relative separation values of Table~\ref{tab:summary_sep}.

Similarly, estimation of the ground truth source locations in the run~7 point cloud introduces an additional systematic position uncertainty.
As the $^{137}$Cs sources in the source-search runs were obscured from the detector operator by cardboard boxes, their true positions are also obscured in the LiDAR point cloud.
The ground truth source locations were therefore estimated by first identifying boxes in the LiDAR point clouds and then using knowledge of source locations within the boxes.
We estimate that this position uncertainty is less than ${\sim}5$~cm, and is predominantly in the $xy$~plane.

Finally, the detector response functions $\eta(\theta, \phi)$ shown in Fig.~\ref{fig:ang_resp_2} are computed using a far-field (parallel) photon beam, and thus will be inaccurate in near-field scenarios where beam divergence is significant (even given the correction in (\ref{eq:regularization})).
Some effort was made to limit near-field effects; in the source-search runs for instance, the distances of closest approach to any ground-truth source ranged from $18$~cm to $40$~cm, with an average of $27$~cm.
Geant4 simulations indicate that effective areas computed at $20$~cm can exceed those computed at a more far-field distance of $200$~cm by ${\gtrsim}10\%$.

We also note that the variation in reconstructed background rates $b$ in Tables~\ref{tab:summary_search} and \ref{tab:summary_sep} may influence reconstructed source activities.
The rate $b = 2.2$~cps in run~8 is expected to be the most accurate background value due to the absence of sources during the measurement.
The underestimation of this background rate in most runs is however a minor difference relative to the ROI counts, and thus we expect the effect on reconstructed activities to be small.

\subsection{Future work}\label{sec:future_work}
We have experimentally demonstrated APSL using a handheld detector system to reconstruct the 3D positions $\vec{r}_s$ and activities $w_s$ of multiple $^{137}$Cs point sources.
Future work may include adapting APSL to identify and reconstruct multiple radionuclides in the same measurement, and to include information from the entire spectrum (rather than just the photopeak) in the quantitative response functions $\eta$.
Proper accounting for near-field effects could also improve the reconstruction quality, though using effective area results as a function of distance could present a significantly larger storage and computational burden.
Various algorithm performance improvements and trade-offs such as parameter tolerances and number of ML-EM iterations could be further explored, with reduced-accuracy but real-time reconstructions perhaps enabling path planning techniques to inform the detector search trajectory in real-time.
Similarly, APSL is amenable to online operation, whereby the reconstruction can be refined as more data is collected, rather than completely recomputed.
Such an online implementation would further reduce the computational burden and help give near-real-time results.
Constraining the addition of sources based on scene data (e.g., the LiDAR point clouds) may also improve localization accuracy.
Finally, we will further explore source-separation capabilities in more constrained scenarios such as straight pass-bys at fixed distances.
Our preliminary experiments indicate that such pass-by measurements typically suffer from low signal-to-noise ratios and large amounts of degeneracy in the solution space.
In tandem, we are exploring fundamental statistical limits on the ability to resolve two gamma-ray point sources (especially in the context of these NG-LAMP measurements), a topic that will be covered in a forthcoming work.
\section{Conclusion}
We have performed experimental demonstrations of Additive Point Source Localization (APSL), a gamma-ray point source reconstruction algorithm.
Using $^{137}$Cs gamma-ray data measured by a handheld detector array, APSL reconstructed the correct number of sources in nearly all search test cases, did so with position and activity errors of ${\sim}20$~cm and $20\%$, respectively, and was able to resolve two $8$~{\textmu}Ci sources separated by distances of ${\gtrsim}75$~cm.
These results offer substantially improved quantification of source localization and activity, and increased image interpretability, over traditional ML-EM methods.
Possible applications enabled by these improvements include retrieval of lost radiological sources and the inspection of declared nuclear facilities, especially if APSL is further developed for near-real-time and online operation.
We anticipate that such scenarios will involve much higher-activity gamma-ray sources, further improving the reconstruction performance of APSL.

\bibliographystyle{IEEEtran}
\bibliography{ref/references}

\end{document}